\documentclass[]{aa}

\usepackage{txfonts}

\usepackage{color}

\begin{document}

\title{Wave mediated angular momentum transport in astrophysical boundary layers}

\author{Marius Hertfelder\thanks{\email{marius.hertfelder@gmail.com}}\inst{1}
        \and
                Wilhelm Kley\inst{1}
        }

   \institute{Institut für Astronomie und Astrophysik, Abt.~Computational Physics, Auf der Morgenstelle 10, 72076 Tübingen, Germany
    }

\titlerunning{Wave mediated AM transport in astrophysical BLs}
\authorrunning{M.~Hertfelder and W.~Kley}

\date{Received xx / Accepted xx}

\abstract
        {
        Disk accretion onto weakly magnetized stars leads to the formation of a boundary layer (BL) where the gas loses its
        excess kinetic energy and settles onto the star. There are still many open questions concerning the BL,  for
        instance the transport of angular momentum (AM) or the vertical structure.
        }
        {
        It is the aim of this work to investigate the AM transport in the BL where the magneto-rotational instability (MRI)
        is not operating owing to the increasing angular velocity $\Omega(r)$ with radius. We will therefore search for an
        appropriate mechanism and examine its efficiency and implications.
        }
        {
        We perform 2D numerical hydrodynamical simulations in a cylindrical coordinate system $(r,\varphi)$ for a thin,
        vertically integrated accretion disk around a young star. We employ a realistic equation of state and include both
        cooling from the disk surfaces and radiation transport in radial and azimuthal direction. The viscosity in the disk
        is treated by the $\alpha$-model; in the BL there is no viscosity term included.
        }
        {
        We find that our setup is unstable to the sonic instability which sets in shortly after the simulations have been
        started. Acoustic waves are generated and traverse the domain, developing weak shocks in the vicinity of the BL.
        Furthermore, the system undergoes recurrent outbursts where the activity in the disk  increases strongly.
        The instability and the waves do not die out for over 2000 orbits.
        }
        {
        There is indeed a purely hydrodynamical mechanism that enables AM transport in the BL. It is efficient and wave
        mediated;   however, this renders it a non-local transport method, which  means that models of a effective local viscosity
        like the $\alpha$-viscosity are probably not applicable in the BL. A variety of further implications of the non-local
        AM transport are discussed.
        }


\maketitle

\section{Introduction}

One ubiquitous phenomenon in astrophysics is the accretion of matter on a central object via an accretion disk. This
process can be observed for a variety of central objects, such as young stars, compact objects like white dwarfs or
neutron stars, and active galactic nuclei. The physics of the accretion disk itself is reasonably well understood.
However, the tiny region where the accretion disk connects to the star is still one of the major  problems.
In this region, called the boundary
layer (BL),  a smooth connection is established between the disk, which rotates nearly at Keplerian frequency, and the star, which
in general rotates much more slowly. It is of great
dynamical and thermal importance since the gas undergoes a supersonic velocity drop in a very confined region of a few
percent of the stellar radius and loses up to one half of the total accretion energy during that process
\citep{1987PhDT.......161K}.

The BL has been studied extensively for over 40 years, both in an analytical \citep[e.g.][]{1992ApJ...399L.163B} and a 
numerical approach \citep[e.g.][]{1987A&A...172..124K}. Simulations involving the BL around a young star have been 
performed by \citet{1996ApJ...461..933K} and \citet{1999ApJ...518..833K}, including full radiation transport. Most of
the work that has been done on the BL \citep[see e.g.][for a brief review of the history]{2013A&A...560A..56H} assumes
that the gas  first slows down in the equatorial plane of the disk and then spreads around the star. This is the
classical picture of the BL. However, there is an alternative theory called the \emph{spreading layer} 
\citep{1999AstL...25..269I,2010AstL...36..848I}, which was initially developed for neutron stars. Within this concept
the gas  first spreads around the star because of  the ram pressure in the BL and then slows down on the whole surface of the star.

The majority of the simulations of the BL start from the premise that the observed angular momentum (hereafter AM) transport
in the BL is driven by local turbulent stresses and the authors consequently adopt a $\alpha$-viscosity model
\citep{1973A&A....24..337S} in their simulations. Sometimes the classical $\alpha$-model is modified slightly in order
to prevent supersonic infall velocities \citep{1997MNRAS.285..239K} or, for instance,  to take into account the radial pressure scale 
height in the BL \citep{1995ApJ...442..337P}. The utilization of a local viscosity model was later  
justified for the disk by the discovery of the magnetorotational instability (MRI) which creates turbulence that acts
like a genuine viscosity on macroscopic scales
\citep{Velikhov_1959,1960PNAS...46..253C,1991ApJ...376..214B,1998RvMP...70....1B,2003ARA&A..41..555B,2008NewAR..51..814B}.
However, as pointed out by \citet{1995MNRAS.277..157G} and \citet{1996MNRAS.281L..21A} and recently shown by
\citet{2012ApJ...751...48P}, if the angular velocity  increases with radius, $\text{d}\,\Omega/\text{d}\, r>0$,
the MRI is effectively damped out and the associated AM transport oscillates about zero. Since this situation clearly
applies for the BL, we do not expect to obtain sufficient AM transport through the MRI. There have been various alternative
transport mechanisms proposed, among them the Kelvin-Helmholtz instability \citep{1978A&A....63..265K}, the baroclinic 
instability \citep{1993ApJ...419..768F}, and the Tayler-Spruit dynamo \citep{1973MNRAS.161..365T,2002A&A...381..923S,2004ApJ...610..977P}, 
none of which have yet been proven to efficiently transport mass and AM in the BL.

In this work, we will focus on the AM transport in the BL through non-axisymmetric instabilities and therefore carry
out 2D  simulations in the midplane of the star-disk system using cylindrical coordinates. Recently, a promising
candidate for the transport has been proposed and investigated in a series of papers. According to this theory, the
steep velocity drop in the BL is prone to the sonic instability \citep{1988MNRAS.231..795G,2012ApJ...752..115B}, 
which is an instability of a supersonic shear layer, much like the 
Papaloizou-Pringle instability \citep{1984MNRAS.208..721P,1987MNRAS.228....1N}. Acoustic waves are excited in the BL
as a consequence of the sonic instability and AM can be transported by these modes in an efficient way. This has been
demonstrated for 2D flows \citep{2012ApJ...760...22B}, 3D flows in cylindrical coordinates \citep{2013ApJ...770...67B},
and even for 3D magnetohydrodynamical flows \citep{2013ApJ...770...68B}. The wave mediated AM transport implies that
this process is intrinsically non-local since the waves can potentially travel a long way before they dissipate and
release the AM to the fluid. Therefore, it is problematic to describe the AM transport in the BL by means of
a local viscosity like the $\alpha$-model. Although the above-mentioned simulations are already
quite sophisticated, the authors make some simplifications that constrain the validity of their models. This is the
point where we step in with this paper and relax three of the simplifications made. First, we make use of a realistic equation of
state instead of an isothermal one and propagate the temperature of the system, as well. We  use full radiative
diffusion in the disk plane and an approximation for the vertical flux, and hence employ a quasi-3D
radiation transport. Second, we use realistic, state-of-the-art initial models like those already published in
\citet{2013A&A...560A..56H} as a starting point for the simulations presented here. Third, mass  constantly enters  the
simulation domain from the outer boundary and thus there is a net mass flow through the disk, as  is the case for a
real accretion disk. Because of the high computational costs the treatment of the radiation brings with it, we perform 
2D simulations and do not yet include magnetic fields. Thus, the next steps will be to extend the 
simulations presented here to 3D and also to include magnetic fields.

The paper  is organized as follows. In Sect.~2, the basic physical model is described. We  give an overview of
the equations used for our simulations and present the assumptions we have made. Furthermore, we briefly review how
the AM transport due to stresses in the fluid can be measured and quantized. Section~3 is devoted to the numerical
method that we employ in our simulations. We discuss the numerical code, the initial and boundary conditions, as well
as the model parameters. In Sect.~4 we start the analysis of the results. We first discuss the sonic instability
and the acoustic modes that are excited in the BL. Then, in Sect.~5, we investigate how these instabilities affect
the physical picture of the BL and look deeply into the AM transport and long-term behavior of our models. We
conclude with Sect.~5.

\section{Physics}\label{sec:physics}

In this section, we present the physical foundations for the simulations we have performed and that are  described
later in this publication. In order to investigate  non-axisymmetric behavior in the disk plane, we utilized the
vertically integrated Navier-Stokes equations. Since the focus of our current work lies on instability
developing in the BL, we describe the flow of a perturbed, compressible fluid and also the derivation of the
resulting Reynolds stresses.

\subsection{Two-dimensional equations for a flat disk}

In order to obtain a set of 2D Navier-Stokes equations including radiation transport, we apply a 
cylindrical coordinate system with coordinates $(r,\varphi,z)$ and integrate the 3D equations over the vertical
coordinate $z$. The star lies in the center of the coordinate system and the midplane of the disk coincides with $z=0$.
If we assume a Gaussian profile for the mass density $\rho$, we can define the 2D 
\emph{surface density} by
\begin{equation}
        \Sigma = \int_{-\infty}^{\infty}\rho\,\text{d}z = \sqrt{2\pi}\,\rho(r,\varphi,0)\,H,
\end{equation}
where we introduce the effective pressure scale height $H$, which is a measure of the vertical extent of the disk. Under 
the assumption of hydrostatic balance and an isothermal equation of state in the $z$-direction, the pressure scale height reads
\begin{equation}
        H = \frac{c_\text{s}}{\Omega_\text{K}},
\end{equation}
with $c_\text{s}$ and $\Omega_\text{K}=\sqrt{GM_\ast/r^3}$ being the midplane soundspeed and the Keplerian angular velocity, respectively.
The inverse of $\Omega_\text{K}(R_\ast)$ is proportional to the period of a Keplerian orbit at the surface of the star,
$P_\ast = 2\pi/\Omega_\text{K}(R_\ast)$.
Derivatives with respect to $z$ are neglected, as is the $z$-component of the velocity, $u_z$. Clearly, this is not
a reasonable assumption if one wants to investigate the overall structure of the BL. However, it is still
justified since we restrict ourselves to the study of non-axisymmetric flow structure in this work.

The vertically integrated continuity equation reads
\begin{gather}
        \frac{\partial\Sigma}{\partial t} + \frac{1}{r}\frac{\partial(r\Sigma u_r)}{\partial r} +
                \frac{1}{r}\frac{\partial(\Sigma u_\varphi)}{\partial \varphi} = 0. \label{eq:mass2D}
\end{gather}
We express the momenta equations in terms of the conservative variables and introduce the radial momentum density 
$s=\Sigma u_r$ and the angular momentum density $h=\Sigma r u_\varphi\equiv\Sigma r^2 \Omega$. These variables reflect the 
physically conserved quantities, as opposed to the primitive variables $u_r$ and $u_\varphi$. The conservation of momentum 
in radial direction is then given by
\begin{multline}
        \frac{\partial s}{\partial t} + \frac{1}{r}\frac{\partial(rs u_r)}{\partial r} +
                \frac{1}{r}\frac{\partial(s u_\varphi)}{\partial \varphi} - \Sigma\frac{u_\varphi^2}{r} = 
                -\frac{\partial p}{\partial r} \\ + \frac{1}{r}\frac{\partial(r\sigma_{rr})}{\partial r} + 
                \frac{1}{r}\frac{\partial(\sigma_{r\varphi})}{\partial \varphi} - \frac{1}{r}\sigma_{\varphi\varphi}
                - \Sigma\frac{\partial \Phi}{\partial r}, \label{eq:radmom2D}
\end{multline}
and the equation in azimuthal direction reads
\begin{multline}
        \frac{\partial h}{\partial t} + \frac{1}{r}\frac{\partial(rh u_r)}{\partial r} +
                \frac{1}{r}\frac{\partial(h u_\varphi)}{\partial \varphi} = -\frac{\partial p}{\partial \varphi} \\
                + \frac{1}{r}\frac{\partial(r^2\sigma_{r\varphi})}{\partial r} + \frac{\partial(\sigma_{\varphi\varphi})}{\partial \varphi}
                - \Sigma\frac{\partial \Phi}{\partial \varphi}. \label{eq:azmom2D}
\end{multline}
Here $p$ denotes the vertically integrated pressure $p = \Sigma R_\text{G}T/\mu$, where $R_\text{G}=
k_\text{B}/m_\text{H}$ with Boltzmann's constant $k_\text{B}$ and the mass of hydrogen $m_\text{H}$, $T$ is the
temperature, and $\mu$ is the mean molecular weight; $\Phi=-GM_\ast/r$ is the gravitational potential ($G$ and $M_\ast$
are the gravitational constant and stellar mass). The components of the vertically integrated viscous stress tensor,
$\sigma_{ij}$, can  be found in e.g. \citet{2002A&A...387..605M}. The equation for the conservation of energy is 
given by
\begin{multline}
        \frac{\partial e}{\partial t} + \frac{1}{r}\frac{\partial(r e u_r)}{\partial r} +
                \frac{1}{r}\frac{\partial(e u_\varphi)}{\partial \varphi} = -p\left[\frac{1}{r}\frac{\partial r u_r}{\partial r}
                + \frac{1}{r}\frac{\partial u_\varphi}{\partial \varphi}\right] \\
                + \sigma_{rr}\frac{\partial u_r}{\partial r} + \sigma_{r\varphi}\left[\frac{1}{r}\frac{\partial u_r}{\partial \varphi}
                + r\frac{\partial \Omega}{\partial r}\right] + \sigma_{\varphi\varphi}\left[\frac{\partial \Omega}{\partial \varphi}
                + \frac{u_r}{r}\right] \\
                - 2\sigma_\text{SB}T_\text{eff}^4 - \frac{H}{r}\left[\frac{\partial (rF_r)}{\partial r} + 
                \frac{\partial F_\varphi}{\partial\varphi} \right],\label{eq:energy2D}
\end{multline}
where $\sigma_\text{SB}$ and $T_\text{eff}$ are the Stefan-Boltzmann constant and the effective temperature, and
the factor $H$ follows from the vertical integration.

The last line of Eq.~(\ref{eq:energy2D}) incorporates the radiative flux in the flux-limited diffusion approximation,
\begin{equation}
        \vec{F} = -K_\text{R}\nabla T\,.\label{eq:radflux}
\end{equation}
Utilizing this approach, we assume that the BL is optically thick both in the radial and vertical directions,
where the inclusion of a flux-limiter also allows for the treatment of optically thin regions.
Equation~(\ref{eq:radflux}) has exactly the same mathematical form as molecular heat conduction in a gas. The effective radiative conductivity
$K_\text{R}$ is
\begin{equation}
        K_\text{R} = \frac{4\lambda a c T^3}{\kappa_\text{R}\rho},
\end{equation}
where $\kappa_\text{R}, a$, and  $c$ are the Rosseland mean opacity, the radiation constant and the speed of light, and
$\lambda$ is the flux limiter \citep{1981ApJ...248..321L,1984JQSRT..31..149L}. We adopt the formulation given by 
\citet{1989A&A...208...98K} for $\lambda$. We assume that the disk locally radiates like a blackbody  of temperature $T_\text{eff}$
and associate the vertical flux with the blackbody luminosity divided by the radiating area, $F_z=\sigma_\text{SB}T_\text{eff}^4$.
The factor 2 in Eq.~(\ref{eq:energy2D}) stems from the fact that the disk has two sides. The vertical flux provides the 
cooling of the disk, while the other parts of the energy equation generate heat through viscous dissipation (second line
in Eq.~\ref{eq:energy2D}) or redistribute it in the disk plane (the terms with $F_r$ and $F_\varphi$). The effective
temperature $T_\text{eff}$ is evaluated according to \citet{1990ApJ...351..632H}, who approximates the optical depth
in the vertical direction of the disk (see Sect.~\ref{sec:Teff}). By approximating the flux in the vertical direction
as well, we employ a quasi-3D radiation transport.

We  also adopt a piecewise prescription for the Rosseland mean opacity, where in each temperature regime $\kappa_\text{R}$ is
given by a power-law approximation, $\kappa_\text{R}=\kappa_0\rho^aT^b$, as described in \citet{1985prpl.conf..981L} and
\citet{1994ApJ...427..987B}. Each of the regimes is characterized by a prevailing mechanism that dominates the opacity
in this region. The parameters $\kappa_0,a$, and $b$ for the various opacity regimes are  listed in e.g. 
\citet{2012A&A...539A..18M}.

\subsection{Viscosity}\label{sec:viscosity}

In the simulations presented here, we strictly distinguish between the BL and the disk with regard to viscosity. As
has already been mentioned in Sect.~1, the AM transport in the disk is mainly driven by turbulent stresses due to the
MRI and can therefore be parametrized with the classic $\alpha$-viscosity approach by \citet{1973A&A....24..337S}. The
effective kinematic viscosity is written as
\begin{equation}
        \nu = \alpha c_\text{s}H        \label{eq:nuSS73}
\end{equation}
in this ansatz, where $\alpha$ is a dimensionless parameter whose value is derived from MRI simulations. The
cause of the AM in the BL and its investigation is, however, the subject of this work. Consequently, we do not make use
of an additional viscosity term in the BL region. The entire AM transport in the BL is therefore due  solely to the Reynolds stresses
exerted by the gas.

From a numerical point of view, the distinction between the BL and the disk is made by searching for the global
maximum of the radial angular velocity profile, $\Omega(r_\text{max}) = \Omega_\text{max}$. This has to be done for
every angle $\varphi$ in order to prevent a suppression of non-axisymmetric perturbations. We can then computationally
realize the viscosity as pointed out above by utilizing the following conditions:
\begin{itemize}
        \item If $r<r_\text{max}$, we set $\alpha=0$ in Eq.~(\ref{eq:nuSS73}) and no viscosity is applied (\emph{BL region}).
        \item If $r>r_\text{max}$, $\alpha$ is set to the default value, $\alpha=0.01$, and the viscous stress tensor
        does not vanish (\emph{disk region}).
 \end{itemize}

\subsection{Reynolds stresses}

Gas flows in general can be characterized by a laminar bulk flow with superimposed perturbations. One typical approach
is to decompose the flow variables $f$ into an averaged or filtered part $\bar{f}$ and a fluctuating or subfiltered,
unresolved part $f'$:
\begin{equation}
        f(\vec{r},t) = \bar{f}(\vec{r},t) + f'(\vec{r},t)\,.
\end{equation}
Within the classical Reynolds approach, the averaging is performed over an ensemble of systems, thus $\bar{f}$ is
an ensemble mean. In the case of a statistically steady or stationary flow field, or when it is homogeneous,
\emph{ergodicity} can be assumed. This means that both a sufficiently long time average over many time
scales and a spatial average over many length scales correspond to the ensemble mean. Often, 
practical flows are inhomogeneous and the ensemble mean is replaced by a time average:
\begin{equation}
        \bar{f}_\tau(\vec{r}, t; \tau) = \frac{1}{\tau} \int_{t-\tau}^t f(\vec{r},t')\,\text{d}t'\,.    \label{eq:time_average_r}
\end{equation}
If, on the other hand, there is a symmetry in at least one direction, as there is in our case, a spatial average is also a 
viable solution.

The actual average process in the Reynolds decomposition is a simple arithmetic mean over space or time 
(see Eq.~\ref{eq:time_average_r}). A more common approach in the case of compressible flows, however, is to use a 
density-weighted Reynolds average for the filtered part of the variables \citep{gatski2013compressibility,peyret1996handbook,1999ApJ...521..650B}. 
These density-weighted, or \emph{Favre} variables are given by a decomposition, which reads
\begin{equation}
        f = \tilde{f} + f''     \quad\text{with}\quad \tilde{f}=\frac{\overline{\rho f}}{\bar{\rho}}\,,
\end{equation}
where the bar denotes a Reynolds average and the tilde a Favre average. 

After performing a decomposition (of any kind) of the flow field, one can then derive the Navier-Stokes equations
for the mean or resolved field. They are called the Reynolds-averaged Navier-Stokes (RANS) equations, or, in the case
of a Favre mean, the Favre-averaged Navier-Stokes (FANS) equations \citep[e.g.][]{gatski2013compressibility,peyret1996handbook}.
Those mean field equations contain additional terms,
however, that are dependent on the fluctuating part of the flow variables. In many engineering applications where 
simulations of the mean field are performed, those parts are unknown and have to be dealt with by using appropriate 
closure relations or models. Here, on the other hand, we are running direct numerical simulations and solve for the 
actual values of the flow field variables. From these simulations, we can compute the additional terms, which appear in 
the averaged equations. 

From the FANS equations, we obtain the Reynolds stresses exerted by the fluctuating motion of the gas. For accretion
disks, the dominant contribution is given by
\begin{equation}
        R_{r\varphi} = \overline{\Sigma u_r u_\varphi} - \overline{\Sigma}\tilde{u}_r\tilde{u}_\varphi\,,       \label{eq:R_1}
\end{equation}
where we have already performed a vertical integration. The quantity $R_{r\varphi}$ appears in the angular momentum equation
(\ref{eq:azmom2D}), where it adds to $\sigma_{r\varphi}$. 
We note that for the second part in Eq.~(\ref{eq:R_1}), it is incorrect to use either the mean values of the laminar
flow (i.e. the initial values, when starting from a relaxed model) or assume $\tilde{u}_r\approx 0$. Our simulations
showed that the instability-dominated state has other mean values than the laminar flow state, and that both terms in 
Eq.~(\ref{eq:R_1}) are approximately of the same order of magnitude. Our simulations also showed that
for the quasi-stationary state both temporal and azimuthal averaging procedures yield the same mean values since
the evolution of the BL is significantly slower than an orbital period.
Therefore, we mostly used azimuthal averaging for the mean values since it is
computationally far less expensive.

The vertically integrated Reynolds stress, which has the dimensions of a 2D pressure, can be expressed by a dimensionless
parameter $\alpha_\text{Re}$ via
\begin{equation}
        \alpha_\text{Re} = \frac{R_{r\varphi}}{\Sigma c_\text{s}^2}.\label{eq:alpha_reynolds}
\end{equation}
Comparing the strength of the Reynolds stresses and the viscous stresses, we can identify 
$R_{r\varphi}=-\sigma_{r\varphi}$ and calculate an effective viscous $\alpha_\nu$ from this equation.
This relation is given by
\begin{equation}
        \alpha_\nu = -\frac{R_{r\varphi}}{\Sigma c_\text{s} \ell r\frac{\partial\Omega}{\partial r}},\label{eq:alpha_visc}
\end{equation}
where $\ell$ is the turbulent scale height in the BL, which is approximated to be the smaller of the disk scale height
$H$ and the radial pressure scale height \citep[see][]{1986MNRAS.220..593P,1995ApJ...442..337P}.

\subsection{Transport of mass and angular momentum}

The mass and AM transport in the disk are two important quantities when we are interested in the
effectivity of the Reynolds stresses in the BL. Both result from the 1D stationary hydrodynamic equations,
where they play the role of eigenvalues. The mass flux is obtained from the continuity equation and reads
\begin{equation}
        \dot{M} = 2\pi r\overline{\Sigma u_r}.  \label{eq:massflux}
\end{equation}
The angular momentum flux follows from the momentum equation in azimuthal direction and is given by
\begin{equation}
        \dot{J} = 2\pi r^2 \overline{\Sigma u_r u_\varphi} - 2\pi r^2\overline{\sigma}_{r\varphi} + 2\pi r^2 R_{r\varphi}.        \label{eq:AMflux}
\end{equation}
Most commonly, the AM flux $\dot{J}$ is expressed as a dimensionless parameter $j$, which is the AM flux normalized to
the advective AM flux at the surface of the star:
\begin{equation}
        j = \frac{\dot{J}}{\dot{J}_\ast},\qquad \dot{J}_\ast = \dot{M}R_\ast^2 \Omega_\text{K}(R_\ast) \label{eq:AMflux2}
.\end{equation}
Either flux is defined such that if mass or AM are traveling outward, $\dot{M}$ and $\dot{J}$ are 
positive. The three terms in Eq.~(\ref{eq:AMflux}) can be identified as the advective AM transport, the viscous transport,  and the 
transport due to Reynolds stresses. Both $\dot{M}$ and $j$ are constant with respect to space and time only if
the system has reached a stationary state, i.e. $\partial/\partial t = 0$. Then, $\dot{M}$ equals the imposed mass
flux that is set as a boundary condition. For non-stationary states, $\dot{M}$ and $j$ represent the local
mass and AM flux at a certain time.

\section{Numerics}

\subsection{General remarks}

We used the code \texttt{FARGO} \citep{2000A&AS..141..165M} with modifications of \citet{2008PhDT.......292B} for the 
simulations presented in this paper. \texttt{FARGO} is a specialized hydrodynamics code for simulations of an
accretion disk in cylindrical coordinates $(r,\varphi)$ that exploits the advantages of the FARGO algorithm 
\citep{2000A&AS..141..165M}. The FARGO algorithm is a special method for the azimuthal transport in differentially
rotating disks that is able to speed up the simulations by up to one order of magnitude and exhibits a lower
numerical viscosity than the usual transport algorithm. The speed-up works best when the velocity perturbations are smaller than the bulk flow in azimuthal direction. Since the average azimuthal velocity $\overline{u}_\varphi(r)$ 
changes significantly in the course of the simulations, the background rotational profile for the FARGO algorithm is updated at each
time step in order to gain full benefit of the method. The algorithm has also been shown to produce valid results  when 
shocks are created in the disk \citep{2000A&AS..141..165M}. The radiative diffusion in the 
flux-limited approximation (Eq.~\ref{eq:radflux}) is solved implicitly using a successive over-relaxation (SOR) method 
\citep[see][]{2013A&A...560A..40M}. The code has been tested and used extensively for a variety of applications. In 
order to verify its suitability for simulations of the BL, we compared it to the 1D BL code described in 
\citet{2013A&A...560A..56H} and found a perfect agreement for the 1D models.

\subsection{Boundary and initial conditions}\label{sec:bcs}

To model the BL with a net mass flow through the disk we have implemented specific boundary conditions in \texttt{FARGO}.
At the outer boundary we impose a mass flux $\dot{M}$ which goes into the disk and whose absolute value can be chosen
as a parameter. We require the azimuthal velocity to be Keplerian at the outer boundary,
$u_\varphi(r_\text{out})=r_\text{out}\Omega_\text{K}(r_\text{out})$. Modeling the inner boundary, i.e. the beginning of the
star, is more difficult. Since we are treating the problem in cylindrical geometry in the disk plane, we cannot 
simultaneously simulate the outer parts of the star. Therefore, we start the simulation domain of all our simulations at
$r_\text{in}=1R_\ast$, where $R_\ast$ is the radius of the chosen star, and set the azimuthal velocity to the velocity of
the rigidly rotating star, $u_\varphi(r_\text{in}) = R_\ast \Omega_\ast$, i.e. we impose a no-slip boundary condition in
$\varphi$-direction. The radial velocity is set to a certain fraction $\mathcal{F}$ of the Keplerian velocity
\begin{equation}
        u_r(r_\text{in}) = -\mathcal{F}r_\text{in}\Omega_\text{K}(r_\text{in})
.\end{equation}
This is obviously an outflow BC for the radial velocity at the inner boundary and it is necessary since the matter
which is inserted in the domain from the outer boundary must be allowed to leave again at the inner boundary. The
factor $\mathcal{F}$ determines how deep the inner boundary $r_\text{in}$ is located inside the star: the smaller
$\mathcal{F}$ is, the deeper the domain ranges into the star. The actual choice of $\mathcal{F}$ is a trade-off
between containing a sufficient part of the stellar envelope in the domain and avoiding  densities and
temperatures that are too high at the inner boundary to prevent the simulations from being slowed down too much. A value of
$\mathcal{F} = 10^{-6}$ is used in our simulations. We have tested different values of $\mathcal{F}$ and found that it
does not have any influence on the initial models apart from shifting the radial profiles by a small amount in $r$.
In particular, it does not have an effect on the mass flux through the inner boundary once the initial model is
in an equilibrium state.
Instead, the density adjusts such that $\dot{M}$ at the inner boundary equals the value at the outer boundary.
Apart from the special boundary treatment mentioned above, we impose zero-gradient or Neuman-type boundary conditions
for the remaining variables, which means that the normal derivative at the boundary vanishes.

The 2D simulations were started from fully relaxed 1D axisymmetric BL models which were  generated using
the same methods as in \citet{2013A&A...560A..56H}. Those simulations use a classical $\alpha$-parametrization for the
viscosity with a uniform value of $\alpha=0.01$ both in the disk and the BL. We employed these models to have realistic 
profiles for the density and temperatures as initial conditions. After the 1D models  reached a stationary state we 
 extended them to 2D input data for \texttt{FARGO} and  ran them again with a small number of azimuthal cells until they were
fully relaxed. Following this \emph{burn-in} procedure we started the actual simulations from
these axisymmetric initial conditions with full resolution in $r$ and $\varphi$. Prior to the start, the radial
velocity was  randomly perturbed by $1\%$ of the Keplerian velocity at the surface of the star. The initial profiles
are shown in Fig.~\ref{fig:vphi_outburst_sequence} below  (dashed lines).

\subsection{Model parameters}\label{sec:sims}

In this paper we focus on the non-axisymmetric instabilities and the AM transport in the BL. We chose the physical
system of a BL around a young star, because the ratio of the mass and the radius of the
star, $M_\ast/R_\ast$, is much smaller than, for instance, for a white dwarf, which we  considered in a previous
paper about the BL. This ratio determines the radial scale height in the BL and the smaller $H$ is, the higher the
resolution must be  in order to resolve the BL correctly. Therefore, the computational costs  increase enormously when one goes to more compact objects. However, we expect what we find here to be the generic case.

The young star we consider in this paper has one solar mass, $M_\ast=M_\sun$, and is three times the radius of the
sun, $R_\ast=3R_\sun$. It has an effective temperature of $T_\ast=3000\,\text{K}$ which yields a luminosity of $3L_\sun$.
The star is not rotating, i.e. $\Omega_\ast = 0$, which is slightly below the observations \citep{1979ApJS...41..743C},
but was chosen for the sake of a large velocity drop. We impose a mass accretion rate of 
$\dot{M}=10^{-6}M_\sun/\text{yr}$, which is a typical value for T Tauri stars such as HL Tau \citep{1993ApJ...418L..71H,1994ApJ...435..821L},
and is two orders of magnitude smaller than the accretion rate FU Orionis stars can reach in outburst phases
\citep{1993prpl.conf..497H}. We took $\alpha=0.01$ for the viscosity parameter in the disk and $\alpha=0$ for the BL
(see Sect.~\ref{sec:viscosity}). We consider the following four simulations for this work.
\begin{itemize}
        \item Simulation A: It has a resolution of $512\times 512$ and the radial domain ranges from $R_\ast$ to $2R_\ast$. This
        simulation features the lowest resolution of all runs performed.
        \item Simulation B: This is the reference simulation, and unless stated otherwise, we always refer to it in the text. It has
        a simulation domain of $[R_\ast,2R_\ast]$ and a resolution of $1024\times 1024$.
        \item Simulation C: Here, the number of grid cells is doubled in each direction  and hence the resolution is
        given by $2048\times 2048$. The simulation time for this resolution is very high and we are not able to study
        the long-term evolution for this run.
        \item Simulation D: In order to ensure that the radial domain in simulations A-C is not too small, we  increased
        it to $[R_\ast,5R_\ast]$. While the first three simulations utilized an arithmetic grid, we  chose a
        logarithmic spacing in radius for this run along with a resolution of $1024\times 1024$.
\end{itemize}

\section{ Boundary layer instability}\label{sec:instability}

In this section, we will summarize the key features of the sonic instability and its associated acoustic waves in the
BL \citep{1988MNRAS.231..795G,2012ApJ...752..115B,2012ApJ...760...22B,2013ApJ...770...67B} and demonstrate that
our models are also prone to the sonic instability.

\subsection{Sonic instability}\label{sec:si}

The sonic instability arises in supersonic shear flows and is a non-local instability similar to the Papaloizou-Pringle
instability \citep{1984MNRAS.208..721P,1987MNRAS.228....1N,1988MNRAS.231..795G}. It is the analogue of the classical
KH instability in the regime of a supersonic drop in the velocity of a compressible, stratified fluid. The sonic
instability has recently been studied in the context of the BL extensively by \citet{2012ApJ...752..115B}. 
% We will summarize some important aspects of their work that also affects our results briefly.

There are two ways by which the sonic instability operates: The first  is the \emph{overreflection} of sound waves from
a critical layer that corresponds to the corotation resonance in a rotating fluid, i.e. the radius where the pattern
frequency of the mode matches the rotation frequency of the bulk flow. This mechanism results in a dispersion
relation that features several sharp peaks and hence discrete modes are clearly favored, especially if the density
contrast across the shear interface is high \citep[see e.g.  panel (n) of Fig.~3 in][]{2012ApJ...752..115B}. Each
mode can be associated with a pseudo-energy, or conserved action (angular momentum), which can leak through the critical layer when
a mode is reflected at the corotation resonance \citep{1987MNRAS.228....1N}. The tunneled wave has the opposite sign of
the action and, since it is a conserved quantity, the reflected wave must undergo an increase in its action in order to
globally compensate for the additional negative action. Thus, the reflected wave is amplified. Furthermore, if  a mode is
trapped such that it performs multiple reflections at the corotation resonance, this amplification mechanism, called
the corotation amplifier \citep[e.g.][]{1976ApJ...205..363M}, eventually leads to instability. 

The other destabilizing mechanism is the \emph{radiation} of energy away from the BL. If a mode inside the BL has a
negative action density, it becomes even more negative through the radiation mechanism and, consequently, is amplified,
which  results in further instability. The dispersion relation of this destabilizing mechanism is a smooth
function of the wave vector. Therefore, a continuum of modes are associated with it. This mechanism dominates the
overreflection for a small density contrast on either side of the interface \citep[see again Fig.~3 of][]{2012ApJ...752..115B}.

Each mechanism, the overreflection and the radiation, is most efficient in the regime of large density contrast or
equal density, respectively, and they operate together. The growth rates of the sonic instability are approximately $1/P_\ast$,
and the dependence of the growth rate on the density contrast is small, despite the different destabilizing mechanisms. 
This further distinguishes the sonic instability from the KH instability. In the numerical simulations of the BL, a
very high numerical resolution is necessary to correctly reproduce the growth rate of the sonic instability. Such
a high resolution is computationally very demanding and costly, especially for investigations of the
long-term evolution of the BL. However, as has been shown by \citet{2012ApJ...752..115B} and  revisited in 
\citet{2012ApJ...760...22B}, a lower resolution merely leads to a slower growth of the instability.

\begin{figure}[t]
  \begin{center}
    %\showthe\columnwidth % Use this to determine the width of the figure.
    \includegraphics[scale=.6]{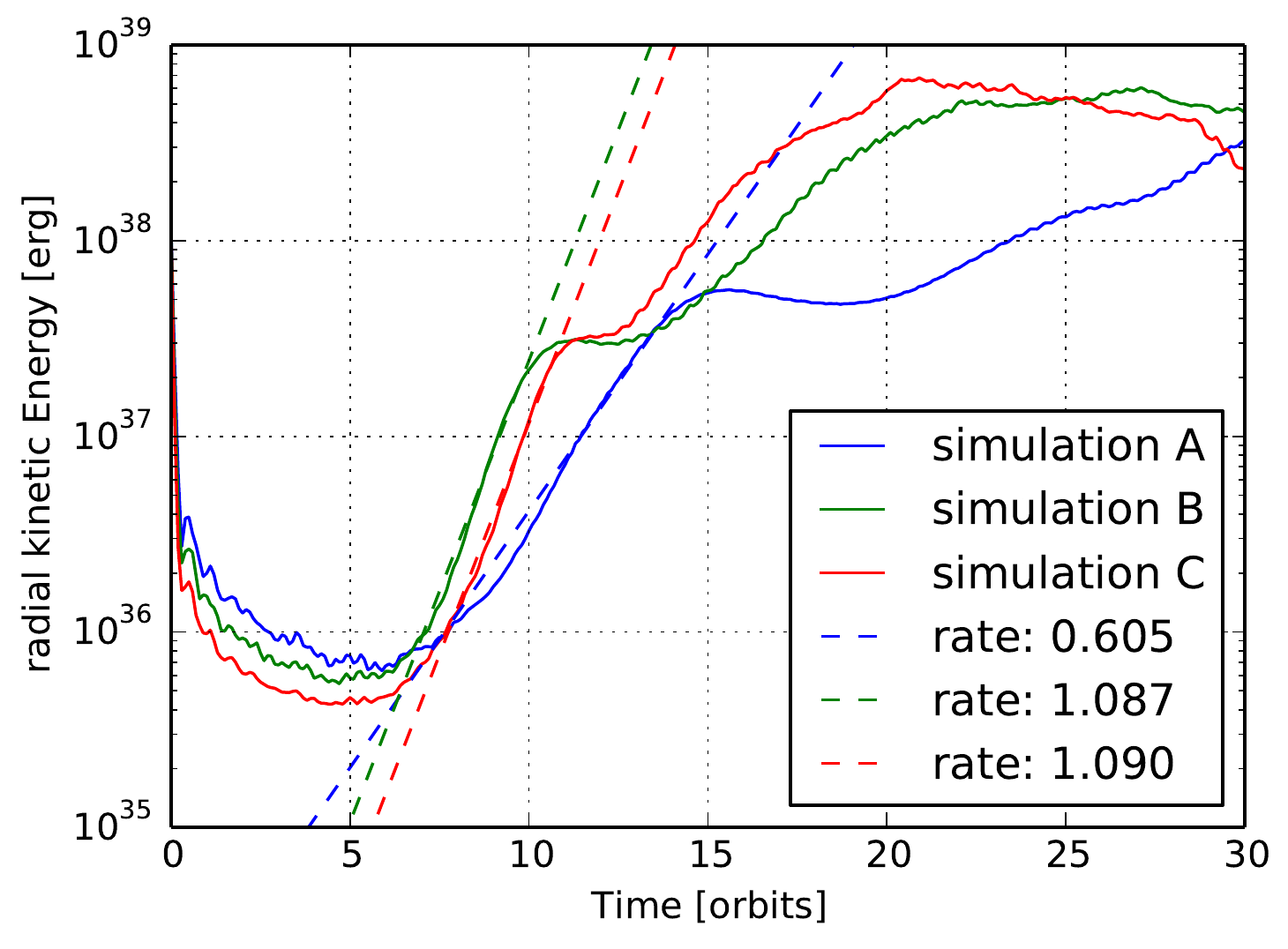}
    \caption{The total radial kinetic energy as a function of the very first orbits for three simulations mentioned in
    Sect.~\ref{sec:sims}, which differ only
    in their spatial resolution. At around orbit 5, the sonic instability starts to operate in all three cases, and
    as a result, the radial kinetic energy  rises rapidly. The growth rates were obtained by applying an exponential
    fit to the first exponential increase. \label{fig:growth_rate_si}}
  \end{center}
\end{figure}

We confirm this point by comparing the growth rates of three simulations with different spatial resolution.
Figure~\ref{fig:growth_rate_si} shows the evolution of the radial kinetic energy when the sonic instability starts
to set in for the three resolutions $512\times512$, $1024\times1024$, and $2048\times2048$. Indeed, the simulation with
the lowest spatial resolution underestimates the growth rate by a factor of $\sim1.8$. However,  the simulations
with $1024^2$ and $2048^2$ cells both yield an almost identical growth rate of $1.09$ per orbit. The shear layer with
positive gradient of $\Omega$ is initially resolved
with approximately $150$ and $300$ cells in radial direction, respectively. In simulation A,
we only have about $75$ cells across the velocity drop. We also verify by Fig.~\ref{fig:growth_rate_si} that the growth
rates of the sonic instability, which are of the order of $1$ per orbit, are actually very high. We will now focus again
on our reference simulation, although all simulations behave basically the same.
After about ten orbits, the rise of the radial kinetic energy is interrupted briefly and then continues with a slightly
different growth rate. The paused increase in the instability and the change in growth rate is due to the change in
density and velocity in the shearing region with time. Since the mass inside the BL is delivered from the disk, the density
in the vicinity of the velocity drop rises. The velocity itself changes as a result of the initiated angular momentum
transport. As has already been mentioned and is elucidated in \citet{2012ApJ...752..115B}, the growth rate of the
sonic instability shows a weak dependence on the density ratio and the modes depend on the slope of the velocity drop, as
well. Thus, a change in the instability mechanism can be observed at this  time.

\begin{figure*}[t]
  \begin{center}
    %\showthe\columnwidth % Use this to determine the width of the figure.
    \includegraphics[width=180mm]{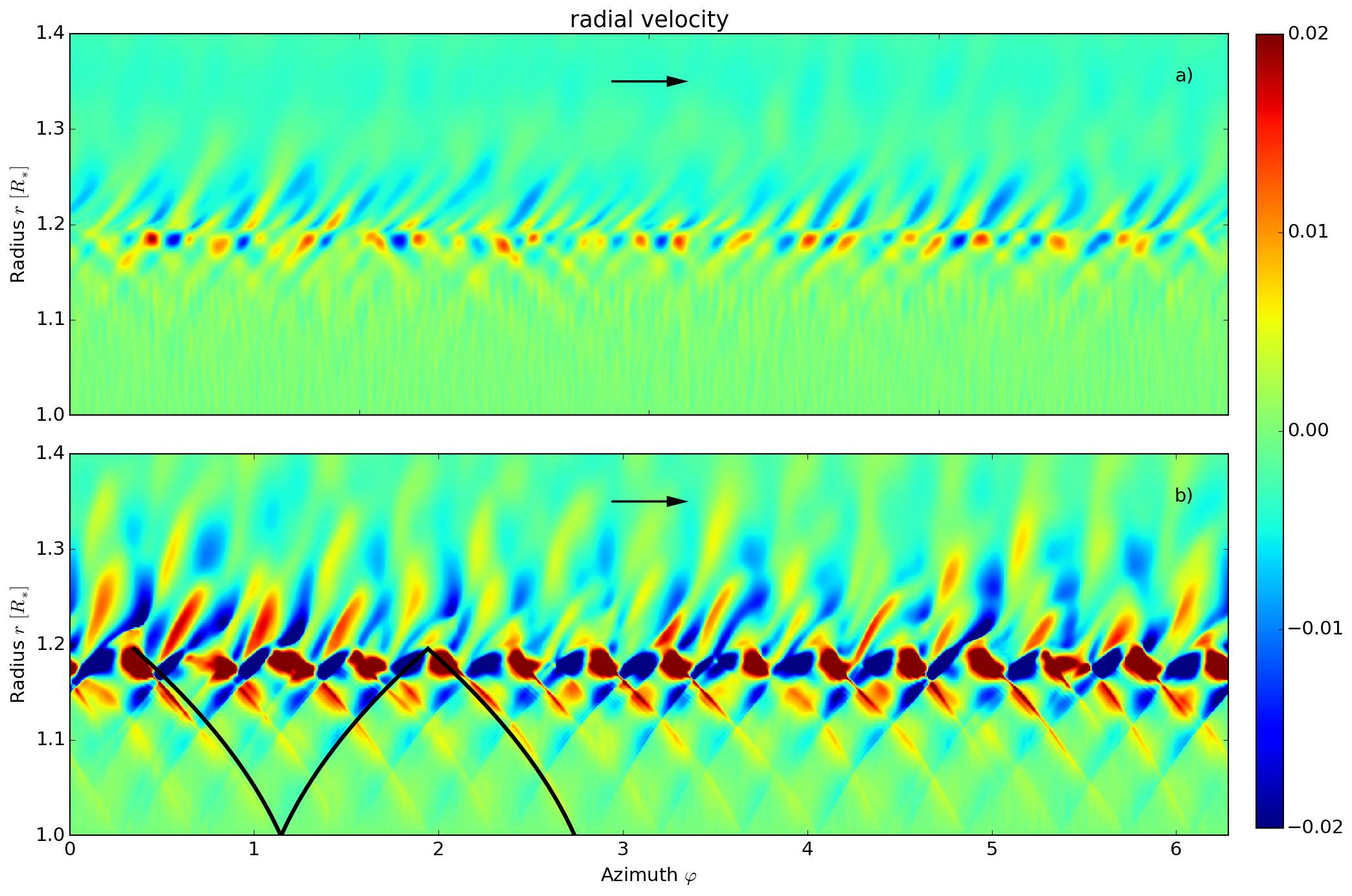}
    \caption{$r$-$\varphi$ plot of the radial velocity in units of $u_\text{K}(R_\ast)$ at $t=5$ (a) and $t=10$ (b) orbits.
    The velocity in the upper panel, a), has been \emph{\emph{multiplied by a factor of 2}} in order to use the same colormap
    as below.
    Both snapshots show the linear stage of the sonic instability and the middle branch in action. The
    sound waves are propagating toward the inner and the outer edge of the domain in an oblique fashion. The three 
    black lines in panel b) illustrate the wavefronts.  
    The arrows denote the direction of the unperturbed azimuthal flow. 
    \label{fig:vrad_si}}
  \end{center}
\end{figure*}

We  now have a closer look at how the sonic instability manifests itself in the radial velocity and consider
Fig.~\ref{fig:vrad_si} for this purpose. The upper panel is a snapshot taken after five orbits and the lower panel shows
$u_r$ at $t=10$ orbits. In both panels, the velocity is normalized to the Keplerian
velocity at the surface of the star. Figure~\ref{fig:vrad_si}a) features a distinct boundary at $r_\text{crit}\approx1.19$, 
which coincides with the inflection point in $\Omega(r)$ and we can clearly see sound waves on either side, propagating
away from the boundary. The pattern speed equals the rotation rate of the unperturbed flow at
the location of the boundary and thus we associate it with the critical layer, or corotation resonance, which we have
already introduced. It is also clearly noticeable that the modes undergo a phase shift across the critical
layer. Both features are a clear signature of the sonic instability and we can compare Fig.~\ref{fig:vrad_si}a) to
Fig.~2 of \citet{2012ApJ...760...22B} and Fig.~5a) of \citet{2012ApJ...752..115B} to verify that our model is indeed
unstable to the sonic instability.

Figure~\ref{fig:vrad_si}b) depicts the radial velocity five orbits later and the
appearance has already noticeably change. The amplitude of the sound waves in the lower panel is at least a factor of 2
larger than that in the upper panel and amounts to a large percentage of the azimuthal velocity. The picture shows a 
distinct pattern of wave front crossings and reflections for $r<r_\text{crit}$ and the critical layer at 
$r_\text{crit}=1.19$ has already begun to smear out at this point in time. When the color scale is saturated we recognize 
three regions where the amplitude of the waves is dropping owing to the wave fronts crossing each other. They lie at 
$r\approx1.16, 1.12$, and $1.07$. Inside the corotation resonance, the waves are reflected near the inner boundary and
at the critical layer, and are thus trapped between the two radii. We have traced three wave fronts with the black lines
to illustrate the pattern that is exactly the overreflection mechanism that we have already elucidated.
The pattern number of the sonic instability (15 for the simulation shown in Fig.~\ref{fig:vrad_si}) does not
depend on details of the inner boundary condition. This has been verified by running the same simulation with identical
perturbations with modified boundary conditions, where all variables are kept at their initial values for all time
(``do-nothing'' BCs). The pattern number is, instead, set by the initial random perturbation. For a different seed of the
initial perturbations the pattern number will be different.
When the trapped waves undergo a reflection at the critical layer, part of the wave action leaks through the critical
layer and has the opposite sign compared to the reflected wave, whose amplitude  grows as a consequence of the global
conservation of wave action. Thus, the whole process is self-sustaining and even grows with time. The leakage of wave
action is also clearly visible in Figure~\ref{fig:vrad_si}b) for $r>r_\text{crit}$. Now the reason for the change in 
the growth rate of the sonic instability after about ten orbits (see Fig.~\ref{fig:growth_rate_si}) becomes apparent:
It is simply the overreflection mechanism setting in, gaining in strength, and growing  dominant.

\subsection{The BL modes}

In Section~\ref{sec:si} we described the linear growth regime of the sonic instability and analyzed its manifestation in 
our models. After only $20-30$ orbits the sonic instability saturates and the system remains in the excited state
with a radial kinetic energy that is approximately three orders of magnitude larger than before the sonic instability
set in. There are still considerable variations in the radial kinetic energy (see Fig.~\ref{fig:Ekin_of_t});  
however, we suggest that they a result of a secondary KH like instability that sets in periodically and dies out shortly after.
This theory is examined thoroughly in Section~\ref{sec:outburst}.

After the sonic instability has saturated we observe waves that are acoustic in nature propagate through the largest part of the domain.
Those modes have been excited by the sonic instability and do not die out for the whole simulation time, which
amounts to over 2000 orbits. Considerable effort has been expended by \citet{2012ApJ...752..115B} and \citet{2012ApJ...760...22B,2013ApJ...770...67B}
in order to analyze these acoustic modes by means of linear stability analysis and derive proper dispersion relations.
Since this undertaking is quite involved we limit ourselves to reviewing those elements that are crucial for the understanding
of our results.

One finds that the acoustic modes of the BL are related to the waves in a plane parallel vortex sheet in the supersonic
regime \citep{2013ApJ...770...67B}. For a discontinuous drop of velocity in the isothermal vortex sheet in absence of
stratification or rotational effects it can be shown that the linearized Euler equations yield three different 
expressions for the dispersion relation $\omega$, which are denoted the \emph{lower, middle}, and \emph{upper branch}
\citep{2012ApJ...752..115B}. The three branches differ in the direction of propagation of the modes on either side
of the interface (velocity drop) and their general appearance is depicted in Fig.~4 of \citet{2013ApJ...770...67B}.
The lower and upper branch are very similar to each other. They both show a region where the waves  propagate
parallel to the interface and a region where the wave vector forms an angle between 0 and $\pi/2$ with respect to
the normal vector of the interface. Transferred to the formalism of our models, the modes of the lower branch 
propagate purely in $\varphi$-direction in the BL and both in $r$ and in $\varphi$ in the disk, where 
$k_r/k_\varphi\gg 1$, however. For the upper branch, the opposite  applies. Modes of the middle branch
propagate in an oblique manner both in the disk and in the BL with the same wave vector. Furthermore, the waves undergo 
a phase shift of $\pi$ at the interface. This holds true for all branches. The dispersion relations of the three branches
of the vortex sheet can then be extended by including radial stratification and the effects of rotation (Coriolis force) 
and one can find approximate dispersion relations for a more realistic case of a BL. This was done in
\citet{2013ApJ...770...67B} for the isothermal BL. It would be interesting to see how a realistic equation of state
and the radiation transport affect the dispersion relations. However, this is beyond the scope of this paper.

\begin{figure}[t]
  \begin{center}
    %\showthe\columnwidth % Use this to determine the width of the figure.
    \includegraphics[width=88mm]{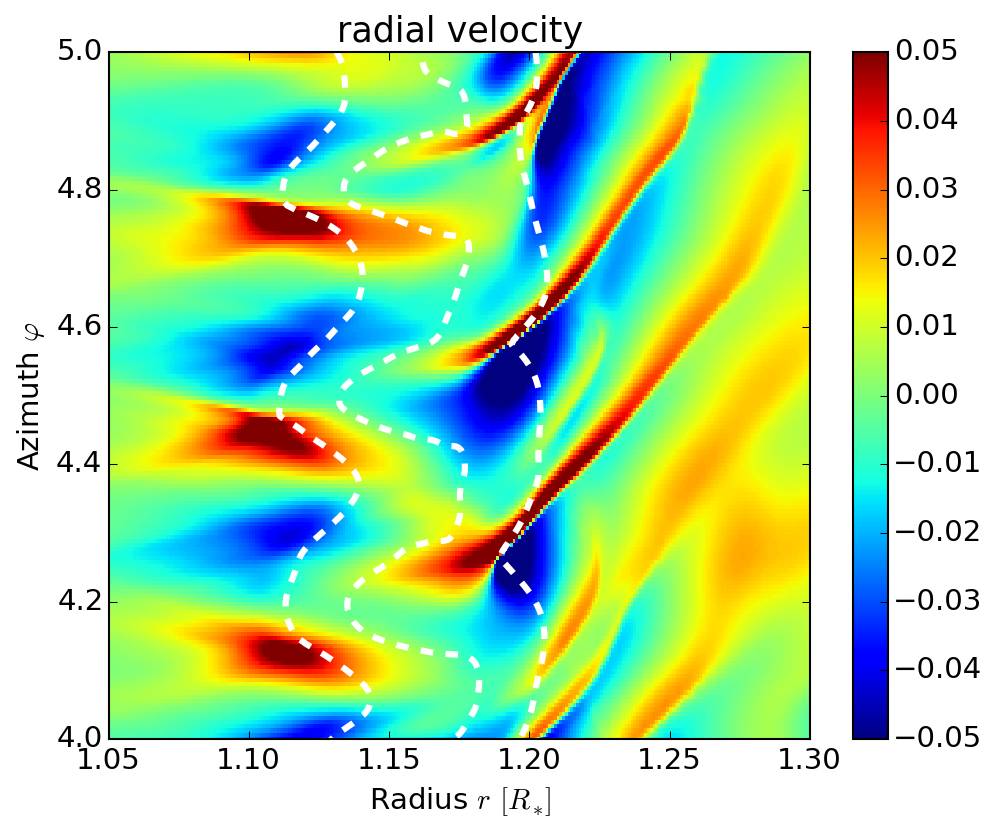}
    \caption{Two-dimensional plot of the radial velocity $u_r$ at $t=45$ orbits. The direction of the unperturbed flow
    is from bottom to top, i.e. in the direction of growing $\varphi$. The dashed white lines are contour lines of the
    azimuthal velocity at the levels $0.15,0.3$, and $0.7$ (from left to right). All velocities are given in units of
    $u_\text{K}(R_\ast)$. The radial velocity develops a pattern with vortices in the BL and shocks at the top of the BL.
    This pattern most likely resembles the lower branch of the acoustic modes discussed in \citet{2013ApJ...770...67B}.
    \label{fig:am_1}}
  \end{center}
\end{figure}

Figure~\ref{fig:am_1} is a $r$-$\varphi$ plot of the radial velocity after 45 orbits. Three white dashed contour lines
of the azimuthal velocity $u_\varphi$ have been overdrawn. All velocities are, as
usual, given in units of the Keplerian velocity at the surface of the star. The image shows a standing wave in azimuthal
direction for radii $\lesssim 1.17$ and a wave that propagates into the disk with wave vector components in both
 $r$- and $\varphi$-directions. Additionally, the amplitude of the wave changes sign at $r\approx 1.17$, i.e. the phase
is shifted by $\pi/2$. This pattern is a clear signature of the lower branch of the acoustic modes, which we
can observe in action here. The contour lines of $u_\varphi$ reveal that the interface between the disk and the BL, or
in other words the region where the velocity starts to deviate substantially from Keplerian, is strongly deformed.
The deformation of the interface is evidently due to the large scale vortices that  reside at the base of the BL (see Fig.~\ref{fig:am_1}). At $r=1.2$ the sound speed amounts to $\sim 8\%$ of the Keplerian velocity, whereas the gas
rotates with $70\% u_\text{K}(R_\ast)$. Therefore, the azimuthal motion is still nearly ten times super sonic and
information about the interface deformation cannot propagate upstream. Hence the oblique shocks, which are clearly 
visible in the radial velocity, have developed. The whole system described above rotates in a prograde way with a pattern speed
of $\Omega_\text{p}=\omega/m$, where $\omega$ is the oscillation frequency in the inertial frame and $m$ the azimuthal
wavenumber. For the episode described here (orbit
45), $\Omega_\text{p}\approx0.3$.

The waves of the lower branch, which we find almost exclusively  in our simulations, are tightly wound with a leading
spiral and we expect $kr/m\gg 1$, where $k$ and $m$ are the radial and azimuthal wave numbers, respectively. Therefore,
one can apply the WKB approximation and assume perturbations of the form
\begin{equation}
        \delta f\propto \exp\left[i\left(kr+m\varphi-\omega t\right)\right].
\end{equation}
A linear stability analysis for a fluid disk with a polytropic equation of state then yields the dispersion relation
\begin{equation}
        \left(\omega - m\Omega\right)^2 = \kappa^2 + k^2 c_\text{s}^2,\label{eq:disp_wkb}
\end{equation}
where $\kappa=\left(\frac{2\Omega}{r}\frac{\text{d}}{\text{d}\,r}(r^2\Omega)\right)^{0.5}$ is the epicyclic frequency
and $\omega/m=\Omega_\text{p}$ \citep{2008gady.book.....B}. Of course, Eq.~(\ref{eq:disp_wkb}) is only an approximation to the dispersion
relation of the waves in our fully radiative disk. However, it will mainly be used to explain the
wave dynamics in the disk. Furthermore, the WKB approximation is no longer valid in the BL where we have seen that
$kr\approx 0$. In the disk we can assume a Keplerian rotation profile, $\Omega\propto r^{-1.5}$, and consequently
the epicyclic frequency is real. Therefore, there is a region around the corotation resonance $\Omega=\Omega_\text{p}$,
flanked by two \emph{Lindblad resonances} $\Omega = \Omega_\text{p} \pm \kappa/m$, where $k$ is imaginary and the
waves are evanescent. Formally, according to Eq.~(\ref{eq:disp_wkb}) the wave number $k$ becomes zero at the Lindblad
resonances. However, the WKB approximation breaks down at these points since the assumption $kr/m\gg 1$ is no longer
valid. Since the whole pattern of waves and interface deformations rotates with a pattern speed $0 < \Omega_p < 1$,
there are two corotation resonances, one in the BL and one in the disk, either of them flanked by two Lindblad resonances.
Assuming the disk rotates with Keplerian frequency, we can easily derive the corotation and Lindblad resonance, which
are then located at $r_\text{cor}=\Omega_\text{p}^{-2/3}$ and $r_\text{Lind}=(\Omega_\text{p} m/(m\pm 1))^{-2/3}$.
The waves or the weak shocks we  discussed earlier launched at the BL now propagate into the disk until they reach
the Lindblad resonances, where they are reflected and propagate back toward the BL. In the BL the waves might be
reflected by Lindblad and corotation resonances as well; however, Eq.~(\ref{eq:disp_wkb}) is not applicable there.
Through this process of consecutive reflections the waves can become trapped between the BL and the disk.

This idea of the trapped modes has been proposed by \citet{2012ApJ...760...22B} and was indeed observed in their 
simulations. In Fig.~\ref{fig:am_1} we do not find such clear evidence for reflected shocks. Reflection of the
wave depends, however, on whether  the modes are absorbed before they reach the evanescent region in the disk. We
observe that the modes in our simulation can only travel until $r\approx 1.5$ before they are substantially damped by
shocks, radiation, or viscous damping due to the $\alpha$-viscosity applied in the disk (see Sect.~\ref{sec:viscosity}).
We have run simulations with larger domains in order to investigate a possible influence of the outer boundary on the
reflection of the waves. These test runs confirm the early damping of the waves observed in the simulations presented
in this work. The effective viscosity in the disk is likely to play a major role in this damping process. The
actual interaction between the waves and the MRI turbulence in the disk can, however, only be investigated in real
MHD simulations.

\begin{figure*}[t]
  \begin{center}
    %\showthe\columnwidth % Use this to determine the width of the figure.
    \includegraphics[width=180mm]{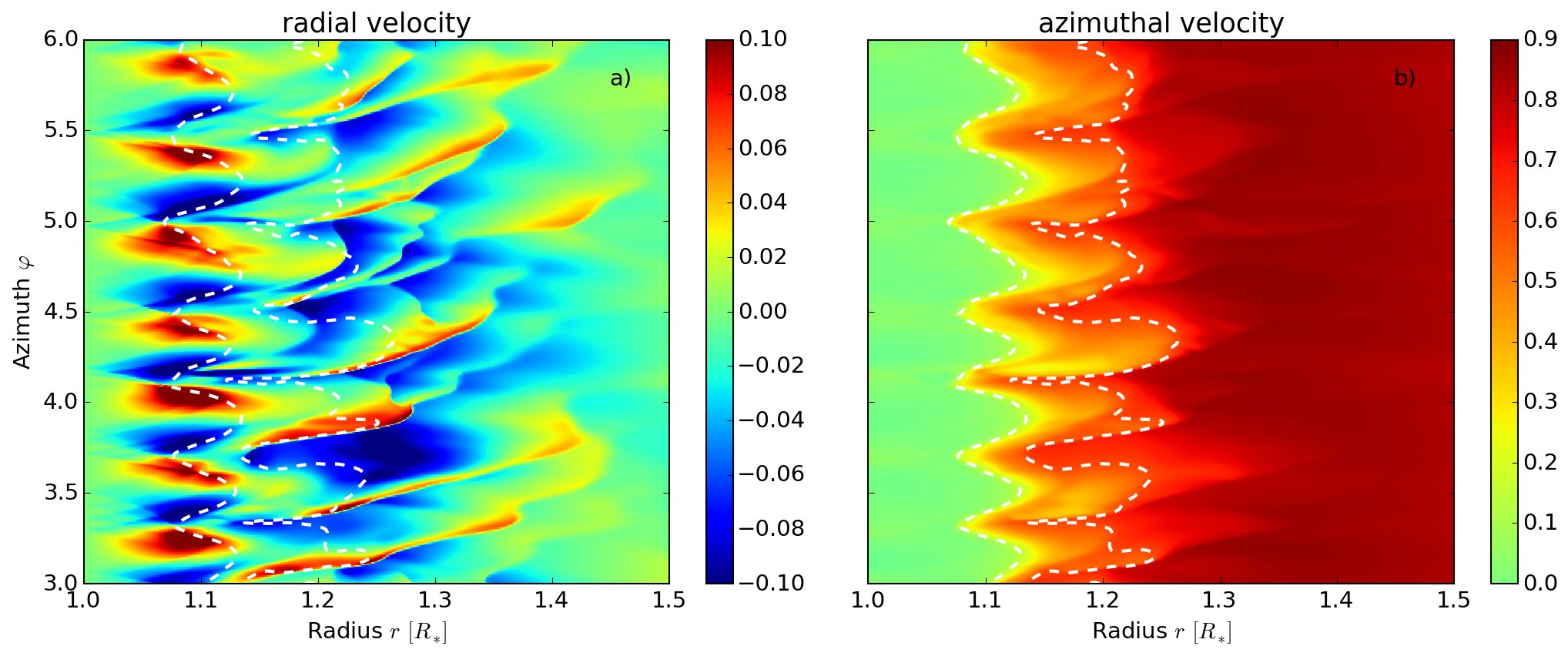}
    \caption{$r$-$\varphi$ plot of the radial velocity (a) and the azimuthal velocity (b) at $t=173$ orbits. Both 
    quantities are given in units of $u_\text{K}(R_\ast)$. The dashed white lines in both plots trace the contour
    of $u_\varphi=0.15$ and $0.6$ (from left to right). The direction of the unperturbed flow is from bottom to top.
    The images have been taken at a high activity state and the strong shocks in $u_r$ and the heavy deformation of
    the interface (see b) are clearly visible.
    \label{fig:am_2}}
  \end{center}
\end{figure*}

In Fig.~\ref{fig:am_2} we plot the velocities $u_r$ and $u_\varphi$ as a function of $(r,\varphi)$ at $t=173$ orbits, a more
energetic state than the one depicted in Fig.~\ref{fig:am_1} (see Fig.~\ref{fig:Ekin_of_t} below). It is immediately
apparent that the state at $t=173$ orbits is far more violent than the situation shown in Fig.~\ref{fig:am_1}. The amplitude of the radial
velocity has increased by a factor of 2 and the shocks have grown more intense and vigorous. While the general pattern
of the radial velocity remains unchanged and still constitutes the lower branch with the vortices at the base of the BL
and the outgoing waves at the top of the BL, the wavefronts are now heavily deformed by  the shocks. These shocks
are a consequence of the strongly distorted interface as can be seen in Fig.~\ref{fig:am_2}b). The lines of constant
$u_\varphi$ in the $r$-$\varphi$ space (the contours) have a wave-like shape with many irregularities and the transition
from zero rotation rate to Keplerian rotation is spread over $\sim 0.2R_\ast$ and has broadened considerably. Though
the radial velocity is more chaotic in Fig.~\ref{fig:am_2}b) than in Fig.~\ref{fig:am_1}, there is still a dominant
mode visible and the azimuthal wave numbers are $m=14$ and $24$, respectively. As we
shall see later, the angular momentum transport by the waves is very efficient in this stage.

\begin{figure}[t]
  \begin{center}
    %\showthe\columnwidth % Use this to determine the width of the figure.
    \includegraphics[width=88mm]{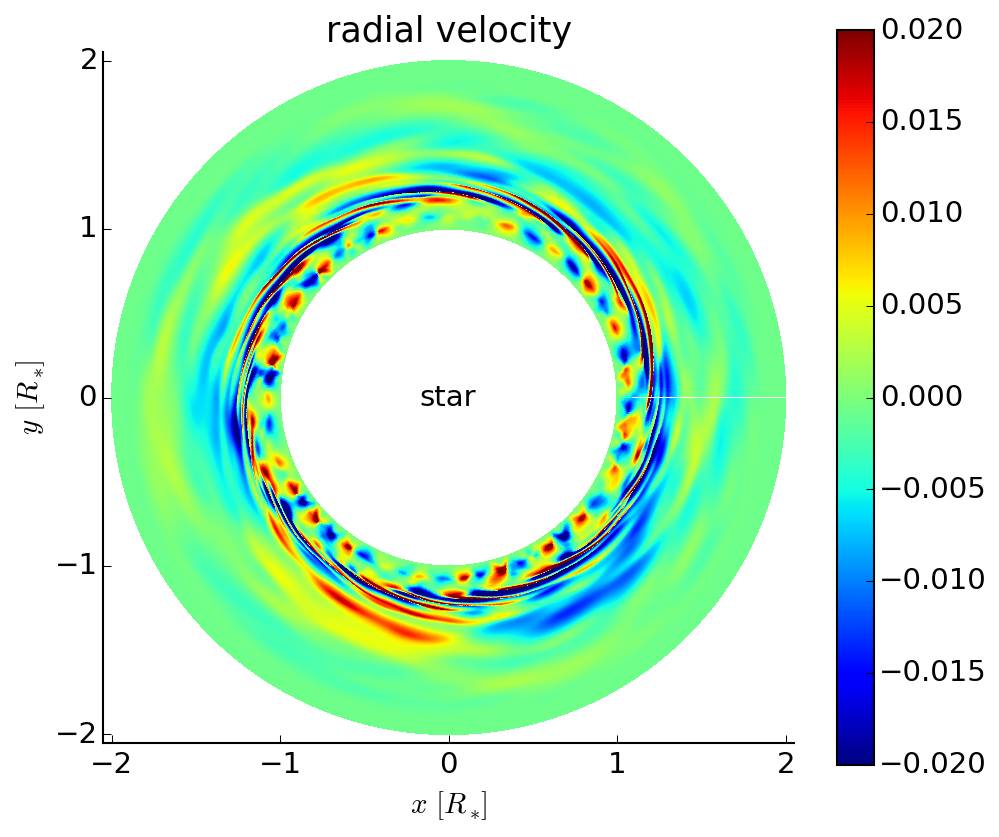}
    \caption{Full-disk plot of $u_r$ at $t=200$ orbits. The direction of the unperturbed flow
    is counterclockwise. The radial velocity is given in units of $u_\text{K}(R_\ast)$. \label{fig:vrad_disc}}
  \end{center}
\end{figure}

For completeness, in Fig.~\ref{fig:vrad_disc} we  show a $r$-$\varphi$-plot of the radial velocity that spans the 
whole simulation domain. The snapshot was taken after 200 orbits while the system was in a quiet state. The pattern
of the lower branch with its vortices at the base of the BL is clearly visible. We can also see that the waves
emerging from the top of the BL are indeed tightly wound and that the spirals are leading. Figure~\ref{fig:vrad_disc}
shows another azimuthal wave with $m=3$ that we had not detected before,  which is still a little faint is this figure.
Interestingly, \citet{2012ApJ...760...22B} also find such a large-scale $m=3$ global mode in their full  $2\pi$
simulations. The wavenumber of the high-frequency pattern at the base of the BL is given by $m=24$

The middle branch is a transient mode and dominates during the linear growth of the sonic instability. It can be observed
in the early stages of the simulation and the oblique wavefronts on either side of the interface are clearly visible in
Fig.~\ref{fig:vrad_si}. Both panel a) and b) show features of the middle branch, yet they are more
distinct in b). The upper branch, on the other hand, might be observed during the late times of the simulations.
We defer the discussion of this issue to Sect.~\ref{sec:long_term}.

\section{The effects of the instabilities}

In   the previous section we  discuss the hydrodynamical instabilities in the BL in detail. Now we  show the
implications of the BL instabilities and the physical results of our simulations of the radiative BL
around a young star.

\subsection{Temporal evolution of the BL}

\begin{figure}[t]
  \begin{center}
    %\showthe\columnwidth % Use this to determine the width of the figure.
    \includegraphics[scale=.6]{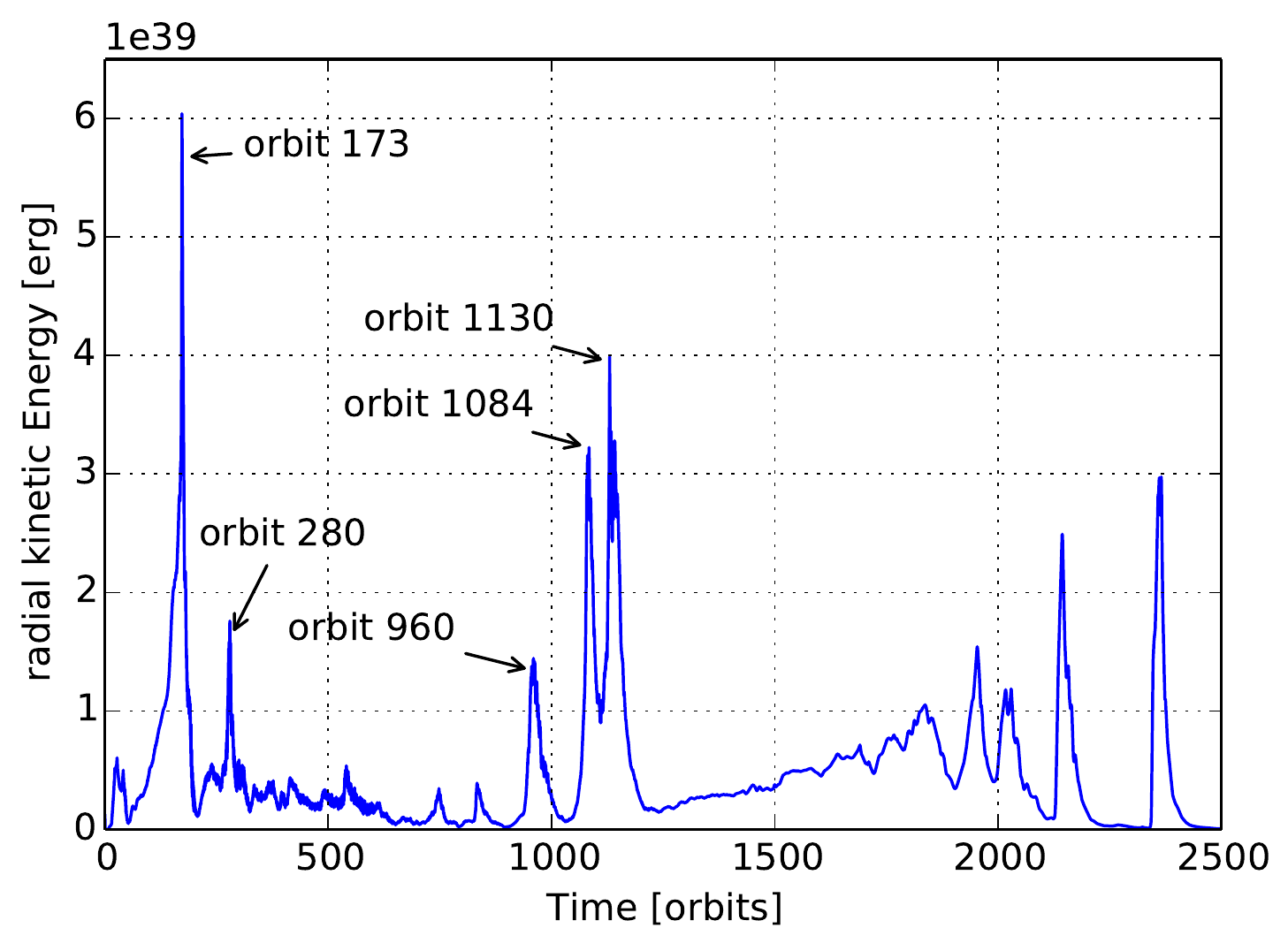}
    \caption{Temporal evolution of the radial kinetic energy of our reference simulation. The time is given in units of
    the Keplerian orbit at $r=R_\ast$, $P_\ast$. The switching between
    quiet and outburst states  described in the text is clearly noticeable. We have marked the exact times for some
    of the outbursts for easier comparison with the text.\label{fig:Ekin_of_t}}
  \end{center}
\end{figure}

Figure \ref{fig:Ekin_of_t} visualizes the total radial kinetic energy,
$E_\text{kin,rad}=\int_\text{disk}\frac{1}{2}\Sigma u_r^2\,\text{d}A$, for which we summed  the radial kinetic energy 
of every cell in the simulation domain. For a given time, the radial kinetic energy reflects the strength of the
instabilities. Therefore, we can monitor the time-dependence
of the instabilities by tracking $E_\text{kin,rad}$. The reference simulation shows a strongly time-dependent behavior that
can be characterized as a recurring series of states of outburst and quiescence. When the kinetic energy reaches a
maximum, the instabilities are vigorous and the interface between the BL and the disk is heavily deformed 
(see Fig.~\ref{fig:am_2}). In the quiescence states, on the other hand, the amplitude of the velocity perturbations 
drops considerably, often by a factor of more than 10, and 
the interface deformation is less strongly pronounced. The strength of the outbursts does not follow a clear pattern.
In the beginning, the outbursts get stronger, with the second outburst at $t\approx 173$ orbits being the 
most powerful. During this outburst the
amplitude of the perturbations in the radial velocity reaches over $20 \%$ of the Keplerian velocity at the surface of 
the star. Thus, the amplitude of the radial velocity becomes comparable to that in
azimuthal direction, or even exceeds it for $r\lesssim 1.1$.
The third outburst at $t\approx 280$, which is less intense than the second one, is followed by a longer period of less activity or
quiescence during which some small peaks suggest the occurrence of very weak micro-outbursts. After more than 500 
orbits, the system starts to go through a series of outbursts again.

\begin{figure*}[t]
  \begin{center}
    %\showthe\columnwidth % Use this to determine the width of the figure.
    \includegraphics[width=180mm]{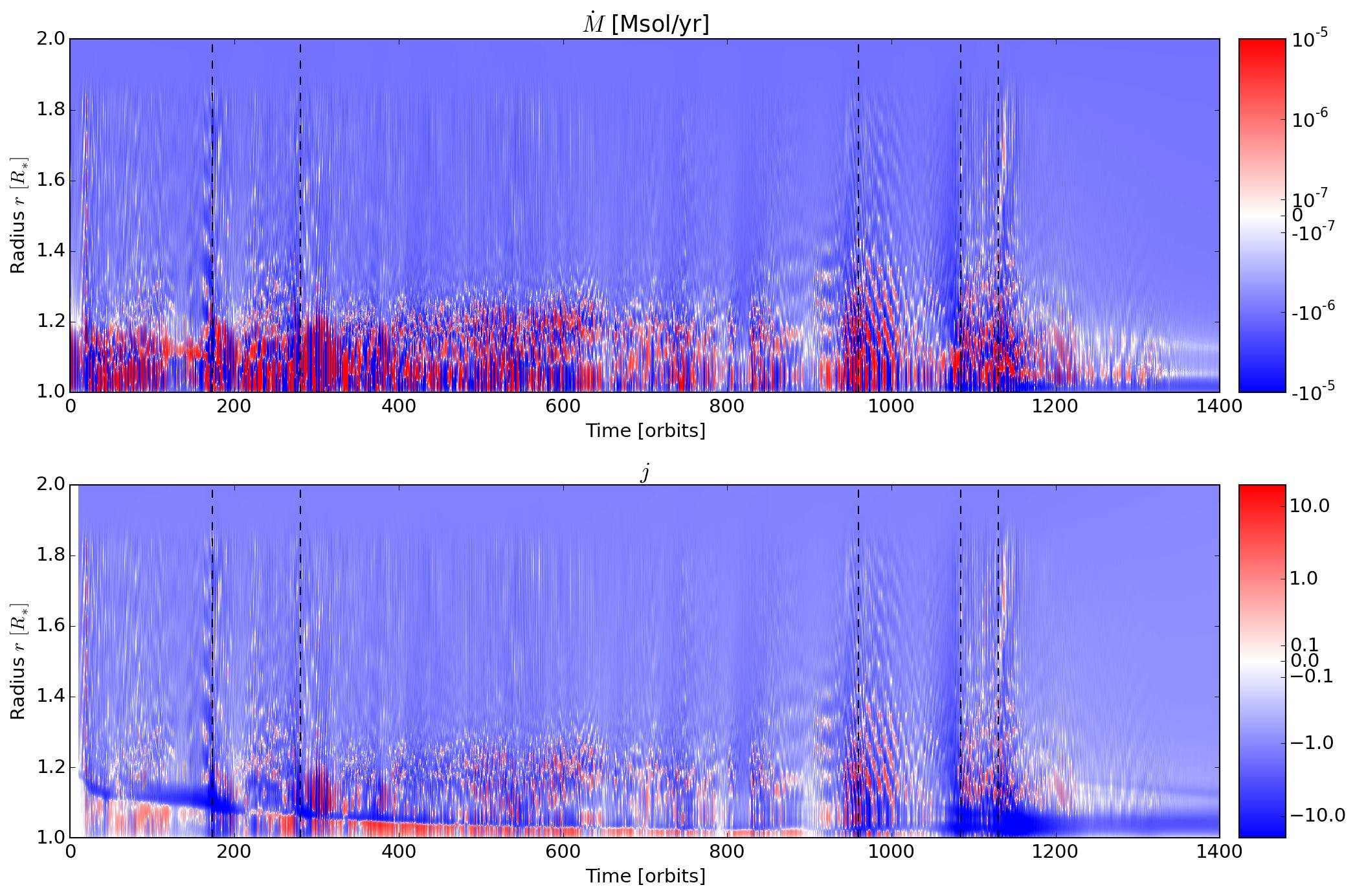}
    \caption{The upper panel shows the temporal evolution of the mass accretion rate in the radial direction, 
    $\dot{M}(r, t)$, in units of solar masses per year. It is obtained from the mass accretion rate per cell 
    (see Eq.~\ref{eq:massflux}) by integrating over the azimuthal direction. The lower panel likewise visualizes the AM 
    flux normalized to the advective AM flux at the stellar surface (see Eq.~\ref{eq:AMflux2}), which is also
    integrated over azimuth. We note that we applied a so-called symlog scaling for the plot, which means that we have a logarithmic scaling for both the 
    positive and the negative values. Only in the vicinity of 0 is the scale linear. A negative value means that mass
    or AM is transported to the center of the disk. The phases of strong activity in the radial velocity 
    (see Fig.~\ref{fig:Ekin_of_t}, marked with the black dashed lines here) correspond to phases of increased mass 
    and AM transport. The diagonal features in the upper panel are a clear signature of waves, which have
    alternating signs of mass transport. This sloshing behavior spreads to the AM flux (lower panel) via the advective
    component of $j$.
    \label{fig:Mdot_Jdot_of_t}}
  \end{center}
\end{figure*}

The high and low instability states mentioned above correspond to episodic phases of high and low activity in mass accretion and
angular momentum transport, as can be seen by comparing Figs.~\ref{fig:Ekin_of_t} and \ref{fig:Mdot_Jdot_of_t}.
Figure 7 depicts the time-dependence of the mass accretion rate, Eq.~(\ref{eq:massflux}), and of the angular
momentum transport, Eq.~(\ref{eq:AMflux}), for all radii inside the computational domain. In phases of outbursts, which
are marked by the vertical dashed lines, the mass flux through the disk is higher than in the quiescence state. 
In certain regions, $\dot{M}$ increases by more than two orders of magnitude. 
In the high activity phases, $\dot{M}$ distinctly rises in the outer parts of the simulation domain as well.
The angular momentum transport, $j$, shows a similar trend. At times of strong instability,  the
transport of angular momentum, which is mainly directed to the center of the disk, also grows remarkably, quite often by a
factor of about 100. 

Unlike in a stationary state where mass and AM are transported to the center of the disk (see Eq.~\ref{eq:AMflux}),
we observe that  the transport of mass does not have a preferred direction in certain regions of the
simulation domain. We consider for this purpose the upper panel of Fig.~\ref{fig:Mdot_Jdot_of_t}.
There is a clear boundary at $r\approx 1.2-1.3 R_\ast$, which is approximately given by the location of the maximum of
$\Omega$ and divides the rather organized, inward directed mass flux in the disk from the alternating transport in the BL, 
where mass is sloshing about back and forth in radius because of  the passage of waves which have alternating
signs of mass transport. This is clearly visible in the form of the diagonal features in the upper panel of Fig.~\ref{fig:Mdot_Jdot_of_t}.
Thus, the direction of the mass transport in the BL is both a function of time and radius, and the variability with time is on
the order of one orbit. After roughly 600 orbits, the strength of the mass transport ceases slightly, only gaining
strength again during the subsequent outburst periods. This development matches that of the radial kinetic energy (Fig.~\ref{fig:Ekin_of_t}) where we can also detect a change in activity starting after 600 orbits. After 1200 orbits
another major change in the mass transport pattern  takes place. The whole system seems to calm down and the sloshing behavior
with its frequent changes in direction  ceases.

\begin{figure}[t]
  \begin{center}
    %\showthe\columnwidth % Use this to determine the width of the figure.
    \includegraphics[scale=.6]{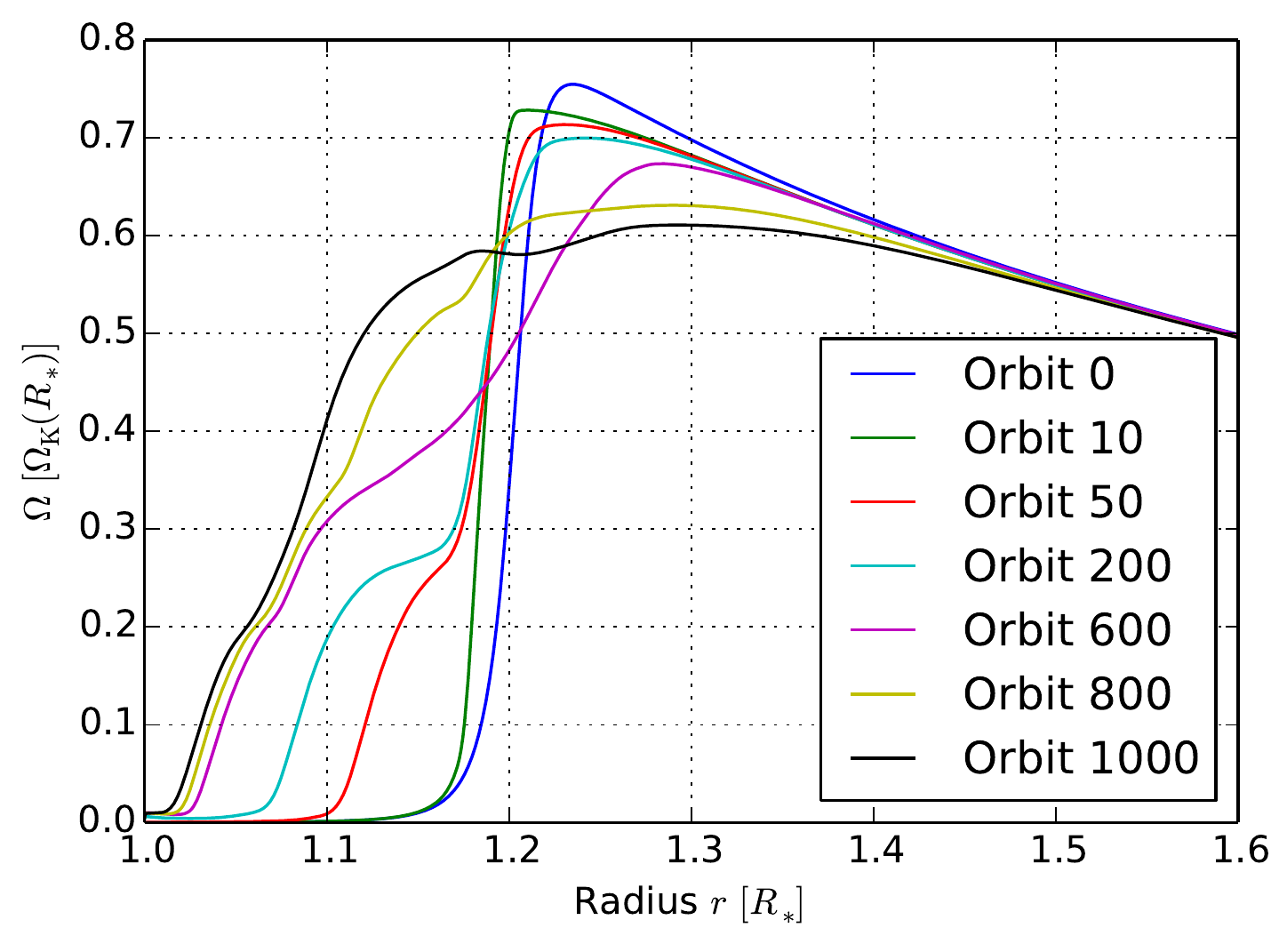}
    \caption{The velocity in azimuthal direction in units of the Keplerian velocity at the stellar surface for seven
    different orbits. The curves demonstrate the spin-up of the gas for $r\lesssim 1.2$ and the spin-down for
    $r\gtrsim 1.2$ over time. \label{fig:AM_of_r}}
  \end{center}
\end{figure}

The AM flux does not feature a variability as intense as the mass flux; however, the diagonal features that have
also been discussed in the context of $\dot{M}$ are still present. An analysis of the individual components of $j$
confirms that these features arise from the advective part of the AM flux, which shows a sloshing behavior similar
to the mass flux. Apart from these features, $j$ assumes mostly
negative values (see the lower panel in Fig.~\ref{fig:Mdot_Jdot_of_t}), meaning that AM is transported inward. 
This evidence agrees with our expectations that in the disk $j$ is negative for the reasons pointed out earlier.
One very distinct feature in the graph of the AM flux is the very pronounced boundary that starts at the beginning of the
simulation at $r$ close to $1.2 R_\ast$ and moves inward up to $r\approx 1.025R_\ast$ during the first 800 orbits.
This line can be characterized by an enhanced AM flux in its near vicinity and a shift in direction from inward
($r<r_\text{line}$) to outward ($r>r_\text{line}$). Figure~\ref{fig:AM_of_r} visualizes the destination of the AM which
is accumulated at this interface as a result of  transport from both sides. It is spent to spin up the gas at $r\lesssim 1.2$. 
Hence, the BL is broadened considerably during the course
of the simulation. The AM needed for the spin-up of the inner layers is extracted from the faster
rotating part at $r\gtrsim 1.2$, which is, consequently, slowed down. By comparing Fig.~\ref{fig:Mdot_Jdot_of_t} and
Fig.~\ref{fig:AM_of_r}, we can deduce that the AM transport is the strongest where the gradient of $\Omega$ is
highest, i.e. where the shearing is strongest.

\begin{figure}[t]
  \begin{center}
    %\showthe\columnwidth % Use this to determine the width of the figure.
    \includegraphics[scale=.6]{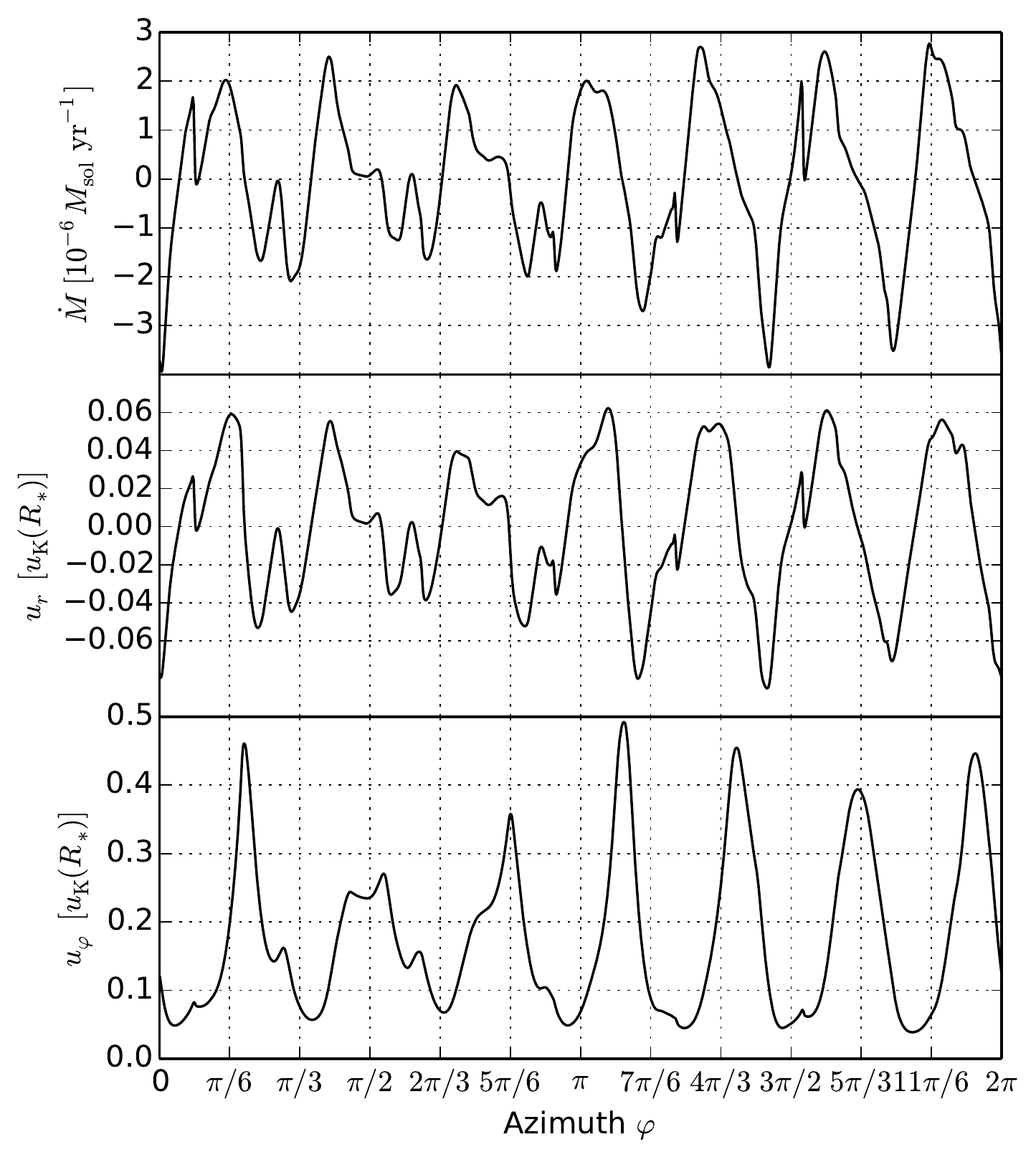}
    \caption{The upper panel depicts the mass accretion rate, the middle panel the radial velocity, and the lower panel the 
    azimuthal velocity. All three quantities are plotted as functions of azimuth at the fixed radius $r=1.12R_\ast$ after 
    145 orbits. The upper two curves are shifted in phase with respect to the lower one such that maxima in $\dot{M}$ (nearly) 
    coincide with minima in $u_\varphi$.\label{fig:anti_correlation}}
  \end{center}
\end{figure}

It can be recognized from Fig.~\ref{fig:Mdot_Jdot_of_t} that AM is mostly transported inward although the mass flux
is also directed outward. Especially when $\dot{M}>0$ and large, one would expect a considerable amount of AM to be 
transported outward along with the matter. Thus, one possibility for the AM flux to still remain negative is that
the Reynolds part of Eq.~(\ref{eq:AMflux}) is negative and larger than the advective part (the viscous part is zero in the BL).
However, an analysis of several time steps showed that very often the advective AM transport is negative and large while the
mass flux is positive and large. This apparent discrepancy between the mass and the AM flux is due to an anti-correlation
between the radial and azimuthal velocity. The mass accretion rate periodically points in- and outward, depending on
azimuth, owing to the vortices of the radial velocity. The sum of $\dot{M}$ over the whole azimuthal domain for a given radius can therefore be positive (meaning 
that mass is transported outward). However, according to Eq.~(\ref{eq:AMflux}), the total advective AM flux can easily
remain negative if $u_\varphi$ is smaller at azimuthal locations of positive $\dot{M}$ than where it is negative. This is
indeed very often the case and we have picked an illustrative example to stress this point (see Fig.~\ref{fig:anti_correlation}).
It is a snapshot of the mass accretion rate, the radial velocity, and the angular velocity at orbit $145$,
where we plotted the dependence on azimuth for $r=1.12R_\ast$. The figure clearly shows that whenever $\dot{M}(\varphi)$ 
is positive and large, $u_\varphi(\varphi)$ is small. The comparison between the radial and the azimuthal
velocity reveals that the anti-correlation between $u_r$ and $u_\varphi$ is such that an extreme of the azimuthal
velocity coincides with a root of the radial one. Therefore, the two patterns that are similar, although not perfectly
equal, are shifted by one quarter of the azimuthal wavelength of the $m=7$ mode at this annulus.

\subsection{Outburst sequence}\label{sec:outburst}

\begin{figure*}[t]
  \begin{center}
    %\showthe\columnwidth % Use this to determine the width of the figure.
    \includegraphics[scale=.6]{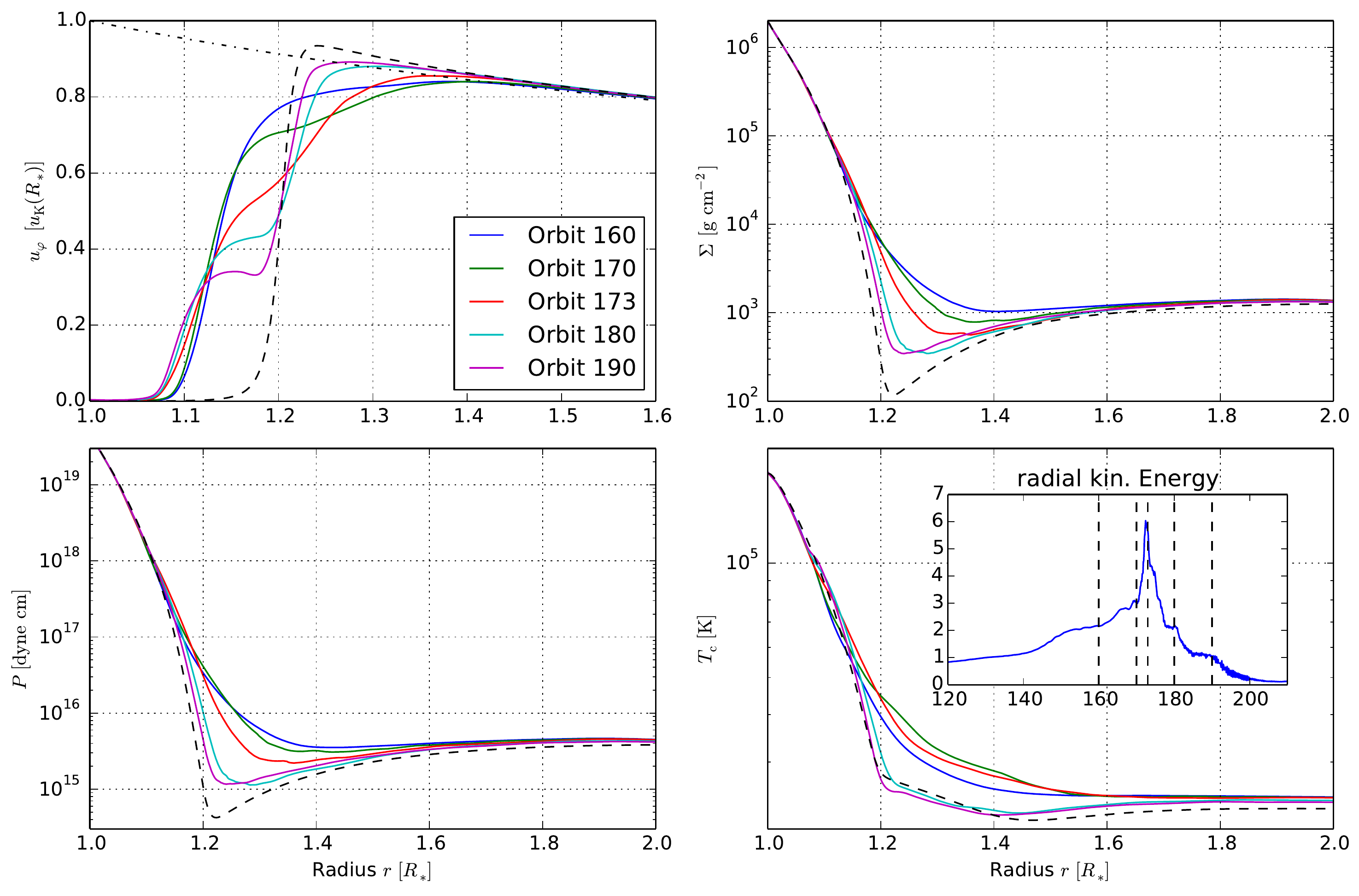}
    \caption{Plots of the azimuthal velocity, the surface density, the pressure, and the midplane temperature for
    five time steps centered around the second outburst. The dashed lines denote the initial conditions of the
    simulations. The dash-dotted line in the upper left panel gives the azimuthal velocity for pure Keplerian
    rotation. In the lower right panel we have included a zoom-in of Fig.~\ref{fig:Ekin_of_t} for better identification
    of the exact location of the plot times.\label{fig:vphi_outburst_sequence}}
  \end{center}
\end{figure*}

We  now take a closer look at the cause of the periodic outbursts and the behavior of the primitive variables in
the disk during such an outburst and we consider Fig.~\ref{fig:vphi_outburst_sequence}
for this purpose. The graphic shows the velocity in azimuthal direction, the surface density, the pressure, and the
temperature in the midplane of the disk for five different times that are centered
around the outburst at $t=173$ orbits and are \emph{not} equidistant. These time steps cover the second outburst in
Fig.~\ref{fig:Ekin_of_t}. This part of the radial kinetic energy evolution is shown as a zoom-in in the lower right
panel of Fig.~\ref{fig:vphi_outburst_sequence}. 

At the beginning of the outburst, the surface density has increased quite strongly, especially in the vicinity of 
$r\approx1.25$ where it is almost two orders of magnitude higher than at the beginning of the simulation. This pile-up of mass is a result of the fact that mass is constantly entering the BL from the outer disk, but the instabilities
in the BL are not yet efficient enough to guide it all through. This situation is universal for all observed outbursts,
which are preceded by a phase of quiescence during which the activity and the mass transport in the BL are not
as strong as in outburst phases. Therefore, the whole process described here using the example of the first outburst
can be equally  applied to the other outbursts. During the accumulation of the
mass just outside the BL, the azimuthal velocity profile is smooth and quasi-stationary. The high amount of mass leads to
a high temperature and hence a high pressure, which means that $u_\varphi$ is remarkably sub-Keplerian in the first half of
the domain since the gas is stabilized by pressure.

After a sufficient amount of mass has accumulated, a runaway process sets in during which $u_\varphi$ and the other 
quantities  change very rapidly. According to Figs.~\ref{fig:Mdot_Jdot_of_t} and \ref{fig:vphi_outburst_sequence}
a great amount of mass is guided through the BL onto the star over such an outburst. The process is initiated outside
the BL at about $r\approx 1.3$. The gas is both accelerated and decelerated in azimuthal direction by the Reynolds stresses in and
near the BL, with the boundary between the spin-up and spin-down region moving inward with time. Since the effect of
the deceleration is much stronger, mass  moves inward. As a consequence of the decreasing density, the pressure
decreases and the gas is even less stabilized and can accelerate its inward movement. After the density is
satisfactorily depleted, the Reynolds stresses decrease and the violent accretion process comes to an end. Then the
density  grows again because of  the lower stresses and the azimuthal velocity is slowly regains its pre-outburst profile.

The  variation of the midplane temperature during the outburst phase is consistent with what we have already discussed for the
other physical quantities. At the beginning, despite the decreasing density, the temperature rises for $r\lesssim 1.6R_\ast$
and even develops a small bump around $r= 1.25R_\ast$. The increase in temperature is due to the dissipation of energy
through the growing Reynolds stresses and indicates the region where the energy release by the waves
is largest. This radius is in agreement with what we expected from the plot of the azimuthal velocity if we also
take into account that the energy is redistributed by radiative diffusion. Subsequently,
the surface density drops more and more and the Reynolds stresses decrease as well. Accordingly, the temperature also
drops, and even drops below the level of the initial state. After enough mass has again piled up around the BL,
the temperature will rise to its pre-outburst level.

The outburst sequence shown here for the second outburst of the reference simulation is exemplary for the general behavior
of outbursts. What we have just discussed might be less clearly visible or even more distinct, depending on the 
strength of the outburst. However, the mechanism is the same throughout. From simulation C, which has a resolution of
$2048\times2048$, we find that the outburst is far more violent than the one we discussed here.
The azimuthal velocity frequently develops an even more pronounced local maximum and minimum for $r < 1.2 R_\ast$  and the depletion of
mass is more effective. Thus, it is possible that the instabilities  become stronger with increasing resolution.
The increasing strength of the outbursts might instead be due to the other modes becoming excited
through the initial perturbations, as  is typical for classical KH instability.

We propose the following scenario for the physical cause of an outburst such as the one described above. The radial
velocity has developed weak shocks at the top of the BL. Every time a fluid element passes through that shock it is
slowed down by these oblique shocks. The cumulative effect of all these shock passages leads to a decrease in the
azimuthal velocity of the gas and creates a wide plateau in the velocity profile (see e.g. orbit 160 in 
Fig.~\ref{fig:vphi_outburst_sequence}). Such a flat region with a weak velocity gradient is prone to the classical
Kelvin-Helmholtz (KH) instability, which sets in and induces the rapid changes in the BL. 
In absence of rotation, it follows from \emph{Rayleigh's stability equation} that a 
necessary condition for instability is the occurrence of an inflection point in the velocity. A stronger form of this
condition was given by \citet{fjortoft1950application}, who showed that the velocity profile not only needs to have an inflection
point, but also has to satisfy the condition $u''(u-u_\text{s})<0$ somewhere in the flow, where $u_\text{s}$ is the 
velocity at the inflection point. However, this requirement is only true for parallel flows. A normal-mode analysis
of the linearized Euler-equations of a rotating incompressible flow with zero mean radial velocity eventually yields 
\begin{equation}
        (\omega+im\Omega)\left(\frac{\partial^2}{\partial r^2}+\frac{1}{r}\frac{\partial}{\partial r}-\frac{m^2}{r^2}\right)\phi - 
        im\frac{1}{r}\frac{\partial Z}{\partial r}\phi = 0 \label{eq:rayleigh}
\end{equation}
for 2D disturbances. Here, $\phi(r)$ is the amplitude of a stream function $\psi'(r,\varphi,t)=\phi(r)
\exp(\omega t+im\varphi)$ such that $u_r'=im\phi/r$ and $u_\varphi'=-\partial\phi/\partial r$, where the prime denotes
perturbed quantities; $Z=1/r\,\partial(ru_\varphi)/\partial r$ is the vorticity. Equation~(\ref{eq:rayleigh})
is the equivalent of Rayleigh's equation for a rotating fluid.  Rayleigh showed that if $\phi=0$ at two
boundaries $R_1$ and $R_2$ enclosing the flow, a necessary condition for instability is that the gradient of $Z$ must
change sign at least once in the interval $R_1<r<R_2$ \citep{2004hyst.book.....D}. We take this analogue of
Rayleigh's inflection point theorem as a motivation to study the vorticity and its derivative, though Eq.~(\ref{eq:rayleigh})
is clearly an idealization of the situation at hand. However, a detailed stability analysis of the RHD equations of a rotating
stratified fluid is beyond the scope of this work.

\begin{figure}[t]
  \begin{center}
    %\showthe\columnwidth % Use this to determine the width of the figure.
    \includegraphics[scale=.6]{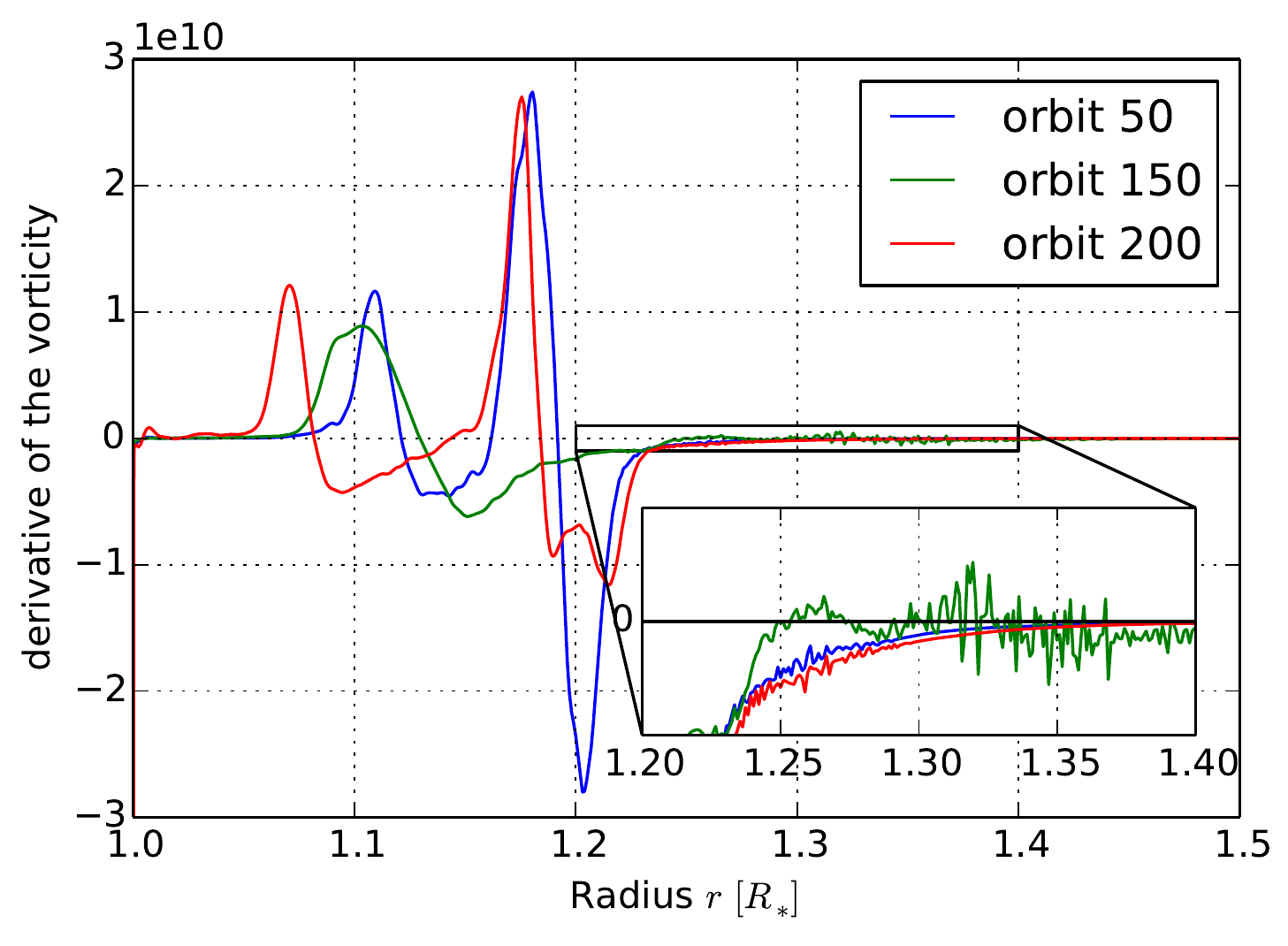}
    \caption{Derivative of the vorticity, $\partial Z/\partial r$, as a function of radius for orbit 50 (quiet state),
    orbit 150 (pre-outburst state), and orbit 200 (post-outburst state). The inset is a zoom-in of the region
    $1.2\leq r \leq 1.4$ where multiple changes of sign are clearly visible for orbit 150.
    \label{fig:vorticity_derivative}}
  \end{center}
\end{figure}

Figure~\ref{fig:vorticity_derivative} displays the derivative of the vorticity, $\partial Z/\partial r$, for three
points in time. At orbit 50, the system is in a state of quiescence (see also Fig.~\ref{fig:Ekin_of_t}). After 150
orbits the second outburst has just started to evolve, and  50 orbits later the system is recovering from
the outburst. While the vorticity derivative of the quiet and post-outburst state does not change sign for $r>1.2 R_\ast$,
it does change several times for orbit 150. The region $r>1.2 R_\ast$ is of interest here because this is where the rotation
profile is flat. There are, of course, multiple changes of sign for $r<1.2R_\ast$ in all curves. However, those points
are associated with the strong shearing regions that we have already discussed in Sect.~\ref{sec:instability}. After analyzing several time steps 
before, during, and after outbursts, we draw the following picture: The consecutive shock passages near the maximum of $\Omega$
cause the flow field to adjust such that multiple changes of sign are created in the derivative of the vorticity. Since
this is a necessary condition for the instability of a rotating, hydrodynamical flow, it is probable that a classical
KH instability develops in the flat regions of the flow. The role, however, that the KH instability plays in the context
of the enhanced accretion and AM transport is not yet entirely clear. It might, on the one hand, act as a trigger for
a mode of the sonic instability which is excited during the outburst and efficiently drives the accretion. Once a
sufficient amount of the mass in the BL has been accreted, the mode can no longer be sustained and the system returns to
quiescence. One candidate for this mode might be the upper branch, which we normally do not observe in our simulations.
A scenario similar to this option was also discussed in \citet{2013ApJ...770...67B}. If, on the other hand, the KH
instability itself is responsible for the enhanced accretion, AM should be transported mainly by turbulent stresses, as
opposed to the wave transport through the mode mentioned above. We did not find clear signs of turbulence during the
outburst, nor did we find evidence of the upper mode. It is, however, plausible that the AM transport is still accomplished
by waves and the KH instability might trigger or enhance a mode of the lower branch.

\begin{figure}[t]
  \begin{center}
    %\showthe\columnwidth % Use this to determine the width of the figure.
    \includegraphics[width=88mm]{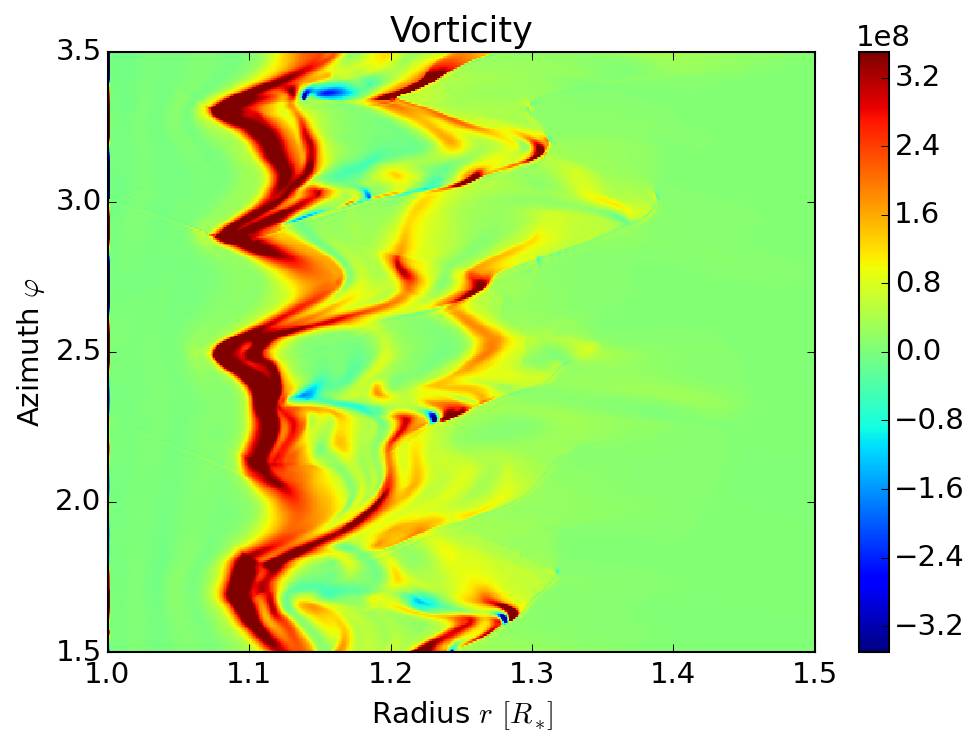}
    \caption{Image of the $z$-component of the vorticity, $\nabla\times\vec{u}$, at the peak of the second outburst
    at $t=172.6$ orbits. The patterns in this image might be indicative of a KH-like instability.
    \label{fig:vorticity_outburst}}
  \end{center}
\end{figure}

We did find evidence for the existence of a KH instability during the outburst. We consider for this purpose Fig.~\ref{fig:vorticity_outburst},
a 2D image of the $z$-component of the vorticity $\nabla\times\vec{u}$ at orbit $172.6$, i.e. at the peak
of the second outburst. It shows patterns
that are commonly associated with the classical KH instability,  for instance, the curly arms and the cat's eyes.
By analyzing simulation C we find that these patterns become more distinct with increasing resolution
\citep[see also Fig.~11 in][]{2012ApJ...760...22B}. The KH features are, however, masked to some extent by the 
waves generated by the main instability. This becomes clear when we investigate the outbursts at later times when the 
main instability is decreasing. In addition, the flat region where we expect KH instability to work 
occurs in the fast rotating part of the disk. Static grid codes suffer from disadvantages when the fluid is boosted
and fluid mixing instabilities can be suppressed \citep{2010MNRAS.401..791S}.

\subsection{Effective temperature}\label{sec:Teff}

\begin{figure}[t]
  \begin{center}
    %\showthe\columnwidth % Use this to determine the width of the figure.
    \includegraphics[scale=.6]{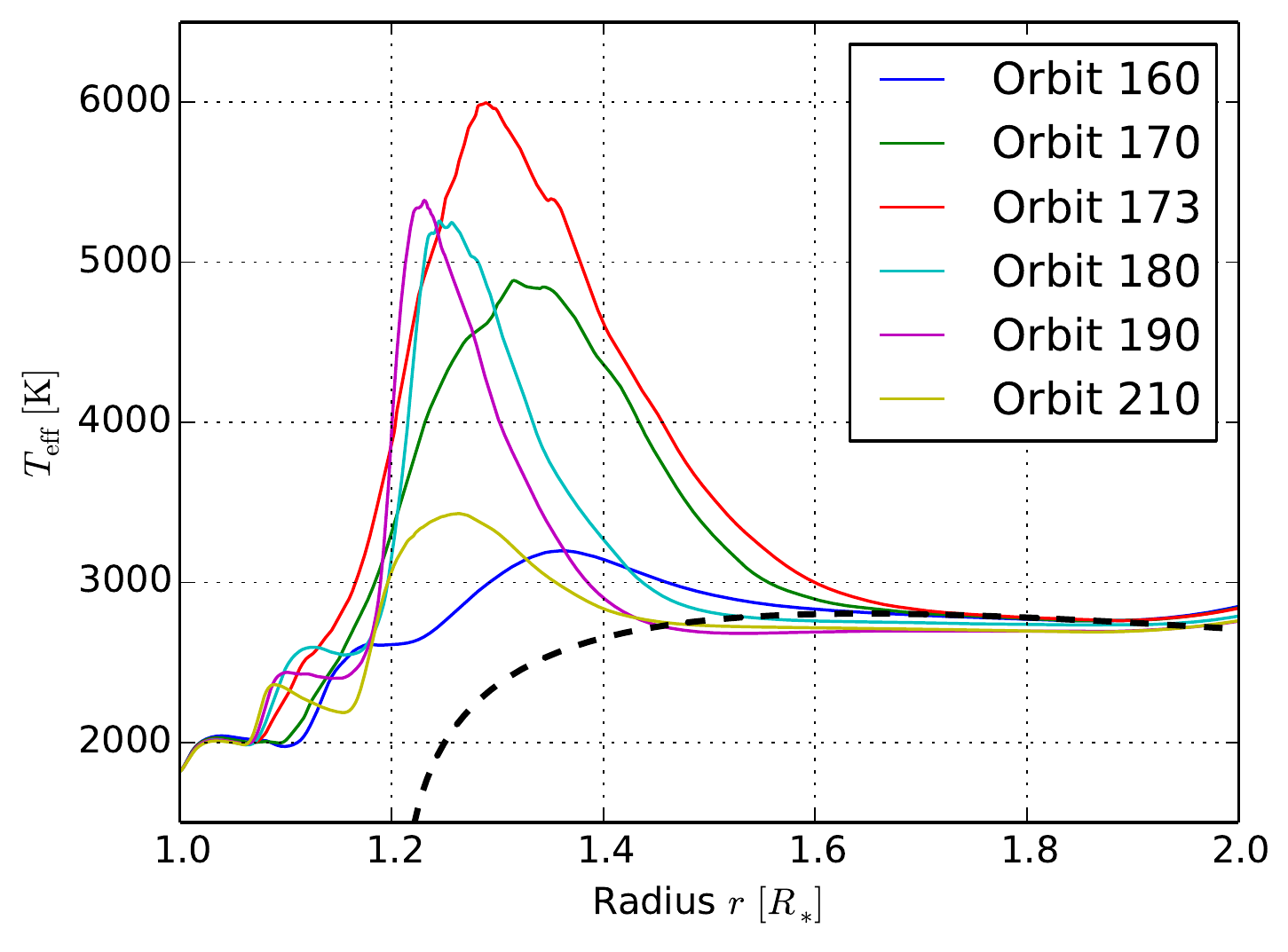}
    \caption{The azimuthally averaged effective temperature of the BL for six time steps centered around the
    second outburst (see Fig.~\ref{fig:Ekin_of_t}). $T_\text{eff}$  increases considerably during the outburst and
    then declines to approximately the pre-outburst level $\sim 30$ orbits after the peak of the outburst. The black
    dashed line corresponds to the surface temperature of the standard solution of \citet{1973A&A....24..337S}.
    \label{fig:t_eff}}
  \end{center}
\end{figure}

We   now investigate the outburst described in Sect.~\ref{sec:outburst} from an observer's point of view and look at the radiation
emitted by the BL and the disk. In Fig.~\ref{fig:vphi_outburst_sequence}
we have exclusively considered physical quantities, which are measured in the equatorial plane of the disk, among
them the midplane temperature $T_\text{c}$. If the medium is optically thick, however, the emergent spectrum is dominated 
by the temperature on the surface of the disk where the vertical optical depth is approximately one, and locally 
resembles the spectrum of a blackbody radiating with temperature $T_\text{eff}$. Since we have no knowledge of the
vertical structure of the disk, we use the approximation of \citet{1990ApJ...351..632H}, which is a generalization of
the gray atmosphere \citep[e.g.][]{1986rpa..book.....R}. The effective temperature can then be calculated with an
approximated optical depth in vertical direction \citep{1992SvAL...18..104S}:
\begin{align}
\begin{split}
        T_\text{eff}^4 & = T_\text{c}^4 / \tau_\text{eff} \\
        \tau_\text{eff} & = \frac{3}{8}\tau_\text{R} + \frac{\sqrt{3}}{4} + \frac{1}{4\tau_\text{P}}. \label{eq:T_eff}
\end{split}
\end{align}
Here, $\tau_\text{R}$ and $\tau_\text{P}$ ($\tau=\kappa\rho H=0.5\kappa\Sigma$) are the Rosseland and the Planck mean 
optical depth, respectively.

Figure~\ref{fig:t_eff} shows the effective temperature of simulation B calculated according to Eq.~(\ref{eq:T_eff}).
At the beginning of the outburst (orbit 160) the effective temperature features only a slight peak in the BL and the temperature
in the BL and in the disk differ at most by 500 Kelvin. The reason why the effective temperature of the
BL and the disk are very similar at the beginning of the outburst episode is the following. The system, which is just on
the verge of an outburst at this point in time, comes out of a preceding state of quiescence. Therefore, the energy
dissipation of the waves is small and the density is rather high owing to the inefficient mass transport. Consequently,
both the midplane temperature and the optical depth in the BL are comparable to the disk and $T_\text{eff}$ is only
slightly larger in the BL. Over the course of the next 13 orbits, however, the system goes through an episode of effective
mass and AM transport, and the midplane temperature increases and the surface density    decreases (see
Fig.~\ref{fig:vphi_outburst_sequence}). Thus, according to Eq.~(\ref{eq:T_eff}), $T_\text{eff}$ 
increases rapidly in the BL yielding a maximum of 6000~K at $r\approx 1.3$, which is twice the temperature in the disk.
Care should be taken when comparing $T_\text{eff}$ in our models with conventional BL models where an $\alpha$-viscosity
has been applied. The latter typically feature a BL that is considerably smaller in radial extent than in the models
presented here. The luminosity of the BL, however, does only depend on the physical parameters of the system and must
be the same in both models ($\frac{1}{2}GM\dot{M}/R_\ast$ for a non-rotating star). Since the luminosity is the
integral of the flux $F=\sigma_\text{SB} T_\text{eff}^4$ over the area of the BL, it follows that the thinner the BL,
the higher the effective temperature must be, and vice versa. The relation between the maximum effective temperature in
the BL $\theta_\text{BL}$ and in the disk, $\theta_\text{d}$, can roughly be estimated by \citep{1974MNRAS.168..603L}
\begin{equation}
        \frac{\theta_\text{BL}}{\theta_\text{d}} = 1.56\left(\frac{\delta}{R_\ast}\right)^{-1/4},       \label{eq:theta}
\end{equation}
where $\delta$ is the width of the \emph{thermal BL} \citep[see e.g.][]{1995MNRAS.272...71R,1995ApJ...442..337P}.
Equation~(\ref{eq:theta}) yields a factor of $\sim 2$ for $\delta=0.3R_\ast$, which is in good agreement with what we observe
in our simulation (see Fig.~\ref{fig:t_eff}).
Also visualized in Fig.~\ref{fig:t_eff} (black dashed line) is the surface temperature according to \citet{1973A&A....24..337S}, which is
given by
\begin{equation}
        T_\text{eff}(r) = \left(\frac{3GM_\ast\dot{M}}{8\pi r^3\sigma_\text{SB}} \left[1 - 
        j\left(\frac{R_\ast}{r}\right)^{1/2}\right]\right)^{1/4}\,      \label{eq:SS_Teff}
,\end{equation}
where $j$ is the normalized AM flux. We took $j=1.1$ for Fig.~\ref{fig:t_eff}, which is the mean normalized AM flux at
orbit 160 measured from the simulation. The effective temperature derived from our simulation agrees very well with
Eq.~(\ref{eq:SS_Teff}) in the disk. In the BL, of course, the analytically derived formula fails to capture the rise
of the effective temperature. 

\begin{figure}[t]
  \begin{center}
    %\showthe\columnwidth % Use this to determine the width of the figure.
    \includegraphics[width=88mm]{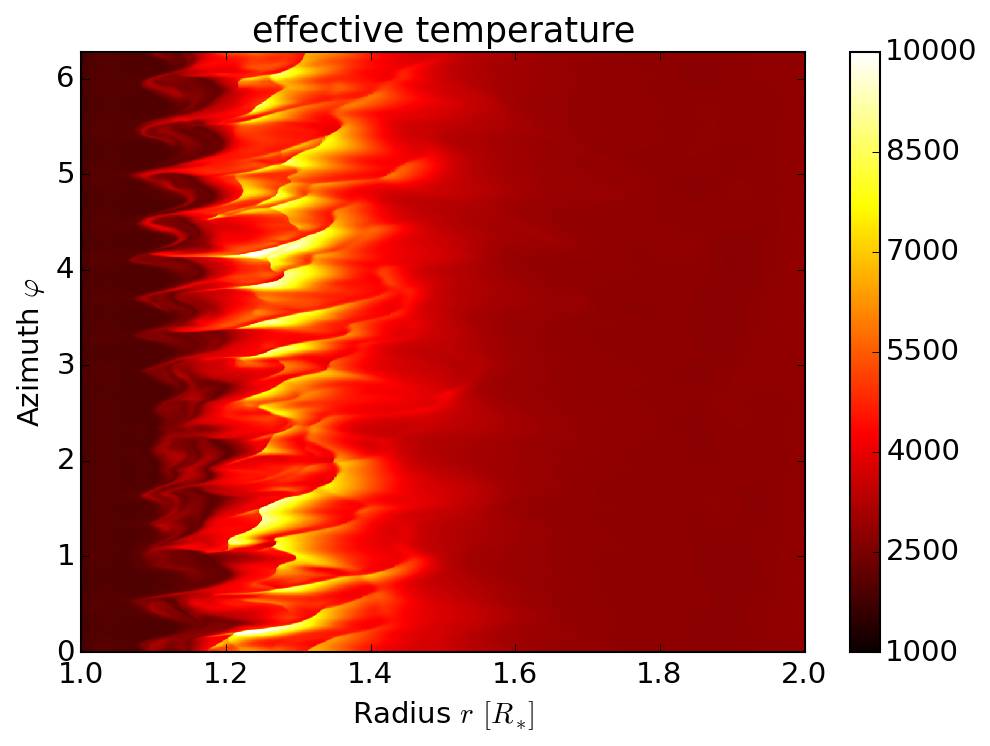}
    \caption{2D plot of the effective temperature of the BL at orbit 173, which marks the peak of the second outburst
    (see Fig.~\ref{fig:Ekin_of_t}). $T_\text{eff}$  strongly varies in azimuthal direction and reaches temperatures
    up to $10\,000$~K. The curly arms visible for $r<1.2$ are indicative of a KH-like instability.
    \label{fig:t_eff_2D}}
  \end{center}
\end{figure}

Another important point for the observational appearance of the BL is the azimuthal dependence of the effective
temperature, which is visualized in Fig.~\ref{fig:t_eff_2D} for orbit number 173. We can  see from this depiction
that $T_\text{eff}$ reaches temperatures of nearly $10\,000$~K in the simulation domain which is high, considering
the case of a young star. There is strong variation of $T_\text{eff}$ in $\varphi$-direction, especially for the
annulus $r\approx 1.3$, which is due to the shocks of the radial velocity that create density perturbations in the
surface density and then affect the vertical optical depth and the effective temperature. The $\varphi$-dependent
pattern of $T_\text{eff}$ rotates with the pattern speed of the acoustic modes, which is close to the Keplerian
velocity, and will express itself in time-dependent variations of the light curve. This is an important point because  many systems show light curve oscillations with periods that are close to the orbital period at the
surface of the star. In cataclysmic variables, for instance, one observes rapid oscillations called dwarf novae
oscillations \citep[DNOs,][]{1972MNRAS.159..101W} and quasi-periodic oscillations \citep[QPOs,][]{1977ApJ...214..144P}.
Both phenomena are not yet fully understood and it is tempting to see a connection to the BL modes on the basis of the
protostar simulations presented here. However, further research is necessary to clarify this point. We also note
that Fig.~\ref{fig:t_eff_2D} shows clear indications of KH instabilities. These features are even more pronounced in
snapshots of $T_\text{eff}$ taken before the peak of the second outburst and reinforce the point that a KH-like instability
is responsible for the frequent outbursts.

\subsection{Angular momentum transport by waves}

In Fig.~\ref{fig:Mdot_Jdot_of_t} we visualized the total AM flux in the disk, which consists of the advective part, the
viscous part in the disk, and the Reynolds part (see Eq.~(\ref{eq:AMflux})). It is, however, of great interest to see
how much the instabilities contribute to the total AM flux and to deduce the efficiency of this transport. As we have
described in Sect.~\ref{sec:instability}, the instability in the BL leads to the development of acoustic waves that are 
launched from the interface and propagate into the disk and the star. As a consequence, AM transport in our models is
dominated by acoustic radiation, i.e. sound waves carrying AM. We do not observe any turbulence in our simulations and
hence rule out turbulent stresses as a major contribution to the AM flux. Unlike the turbulent stresses, the AM
transport by waves is a non-local process. Waves can carry AM of either sign from one point in the domain, e.g. the
interface in the BL where they are launched to another point in the disk where they deposit the AM into the fluid
through wave dissipation. Thus, the  long-range exchange of angular momentum between regions that are otherwise 
uncorrelated is possible through the wave transport. This is a major difference from the $\alpha$-models which yield a
local effective viscosity due to a local AM exchange phenomenon, such as turbulence. Therefore, it is questionable whether the
processes in the BL can be described by a local prescription like the $\alpha$-model.

\begin{figure*}[t]
  \begin{center}
    %\showthe\columnwidth % Use this to determine the width of the figure.
    \includegraphics[width=180mm]{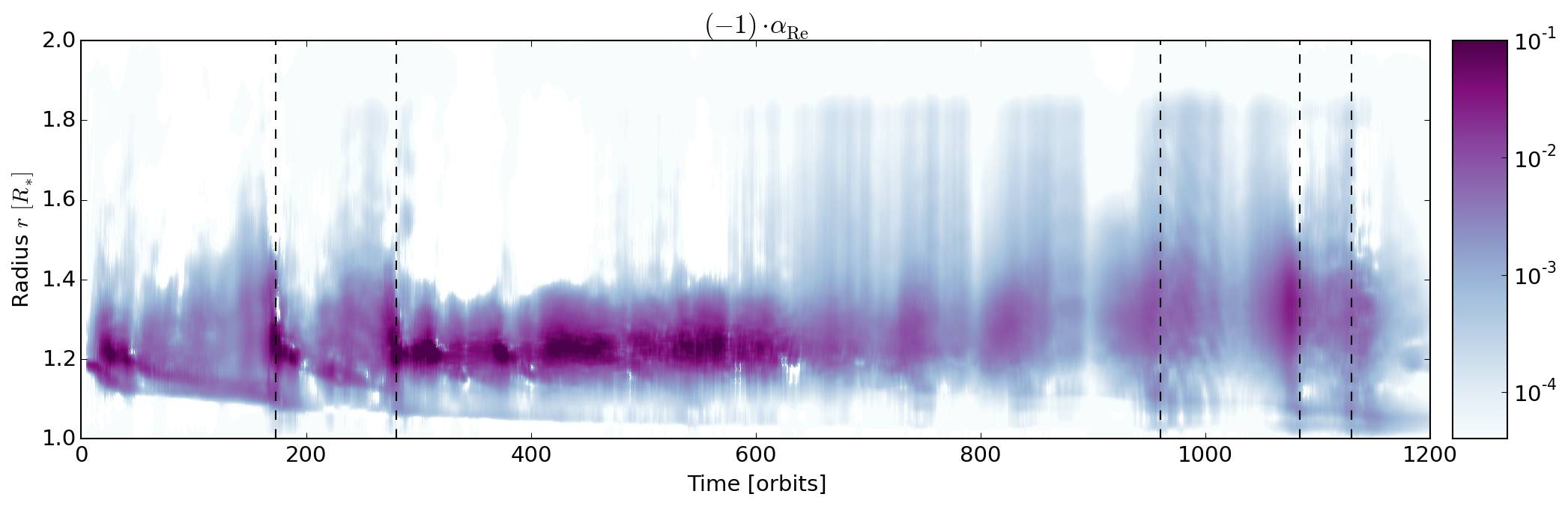}
    \caption{Time-radius image of $\alpha_\text{Re}$ spanning several hundred orbits. We note that the values of
    $\alpha_\text{Re}$ have been multiplied by $(-1)$ in order to allow for the logarithmic scale. For each point in
    time the values have been time average over ten orbits. The vertical dashed
    lines denote the locations of the outbursts mentioned earlier in the text.
    \label{fig:alpha_of_t}}
  \end{center}
\end{figure*}

We  now analyze the efficiency of the wave mediated AM transport and consider Fig.~\ref{fig:alpha_of_t} for this
purpose. It shows the dimensionless Reynolds stresses $\alpha_\text{Re}$, which has been calculated according to
Eq.~(\ref{eq:alpha_reynolds}), as a function of radius and time. We can apply this description to the AM transport 
since waves transport angular momentum exclusively by stresses. The Reynolds stress is evidently very strong with
peak values up to the order of unity and it is largest at the top of the BL, at around $r=1.2$, which   is very interesting
since in that region $\Omega$ reaches its maximum. Thus, our simulations reveal that the stresses are largest around
the very point where they would vanish according to a conventional $\alpha$-viscosity model. Strictly speaking, this 
renders a local prescription that depends on the shear of the angular velocity inapplicable for the BL. 
Figure~\ref{fig:alpha_of_t} also shows that the AM transport due to the waves includes a large part of the domain,
$1.05\lesssim r\lesssim 1.4$, and spreads out farther into the disk during outbursts. This reinforces the point that
the AM transport ultimately generated by the shear in the BL is not confined to the BL, but in fact ranges over long
distances. As we saw earlier, the activity  rises during outbursts and therefore $\alpha_\text{Re}$ is
enhanced during these periods as well. The value of  $\alpha_\text{Re}$ is almost exclusively negative, meaning that angular
momentum is transported toward the star. This behavior is characteristic of the lower branch of the acoustic modes
\citep{2013ApJ...770...67B}. We refrain from presenting plots of the parameter $\alpha_\nu$ that can be calculated
according to Eq.~(\ref{eq:alpha_visc}) because it is artificial to boil the non-local AM transport down to
a local parameter. However, for the sake of completeness, it should be mentioned that Eq.~(\ref{eq:alpha_visc}) yields
values on the order of $10^{-3}-1$ for our models.

\begin{figure}[t]
  \begin{center}
    %\showthe\columnwidth % Use this to determine the width of the figure.
    \includegraphics[scale=.6]{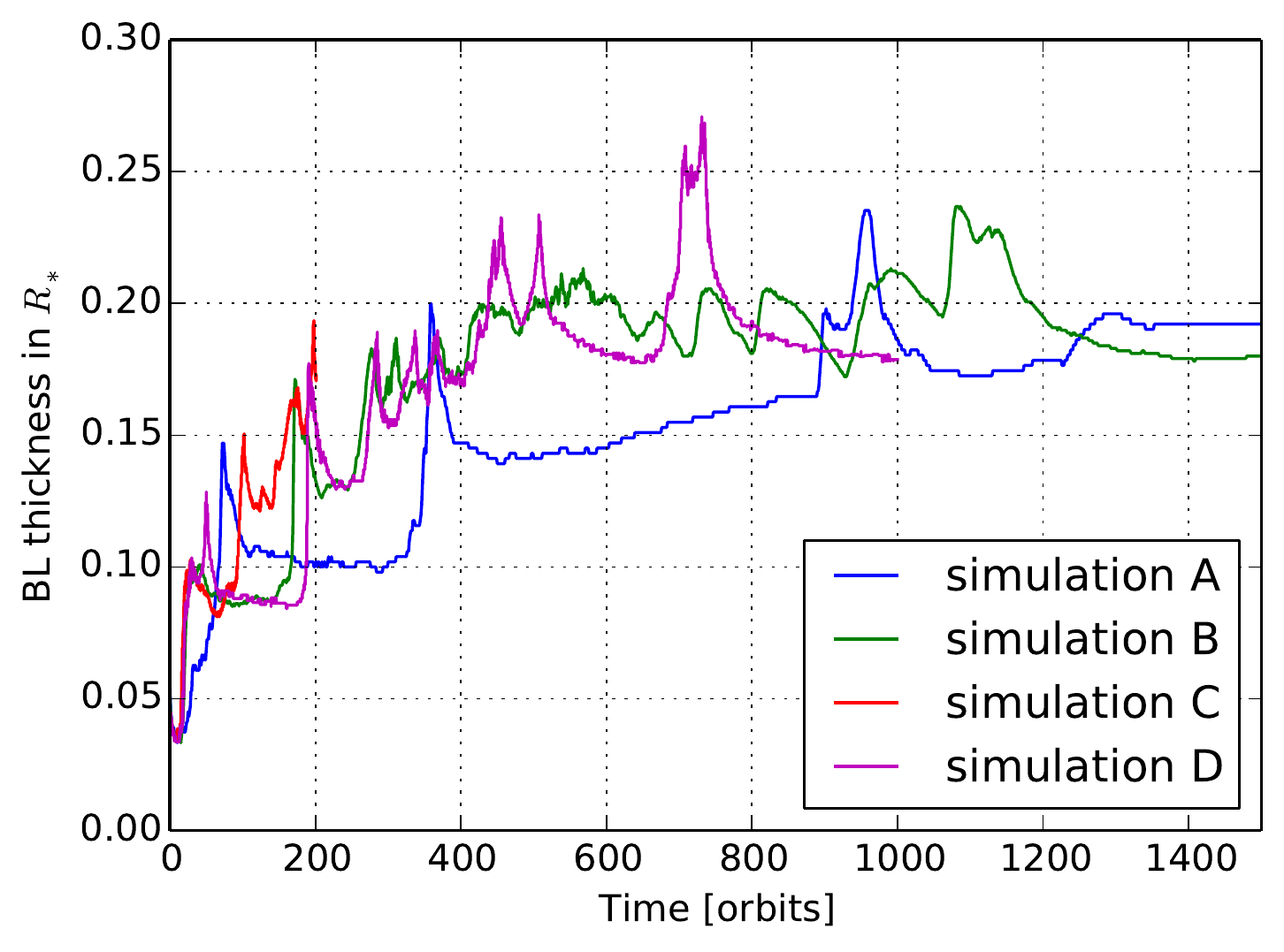}
    \caption{The width of the BL as a function of time for the four simulations mentioned in Sect.~\ref{sec:sims}. The 
    width is calculated from the radii of the region where $0.02<u_\varphi/u_\text{K}<0.8$. All four simulation settle
    down to a BL width of approximately 20 percent of the stellar radius. Simulation A approaches this value more
    slowly and in a slightly different manner, which is due to the very low resolution of this particular simulation
    ($512\times 512$).
    \label{fig:width_bl}}
  \end{center}
\end{figure}

Finally, we investigate how the strong AM transport in and around the BL affects the width of the region where the
azimuthal velocity rises from stellar rotation to the Keplerian rotation in the disk. Figure~\ref{fig:width_bl} depicts
the BL thickness as a function of time for all four simulations. The thickness of the BL is defined to be the region in which
\begin{equation}
        0.02 < \overline{u_\varphi}(r)/u_\text{K}(r) < 0.8.
\end{equation}
The width of the BL increases very rapidly and grows in size by a factor of 2  at the beginning of the simulation, i.e. 
when the sonic instability has grown to establish an efficient AM transport. After the initial fast growth the width
of the BL increases on average linearly over the course of approximately 500 orbits until it settles down at a value of
$\sim0.2R_\ast$. We have run additional simulations in order to test the robustness of the BL thickness of
$20\,\%$ of the stellar radius. In these simulations, the initial distance of the velocity drop from the inner edge of
the simulation domain has been extended by altering both the parameter $\mathcal{F}$ (see Sect.~\ref{sec:bcs}) and the point where the
computational domain begins. We thereby test whether the thickness is affected by  the BL running up against
the inner edge of the domain. However, these runs are in perfect agreement with the simulations presented here and saturate at
a BL width of $\sim 0.2 R_\ast$ as well.

There are several spikes in the temporal evolution of the BL width for all three simulations that are
clearly associated with the frequent outbursts. During the high activity state the BL broadens considerably in a very
short period of time and typically resides at a larger width in the aftermath of the outburst. Because of the high
resolution and the resulting long simulation times we can follow the highly resolved simulation only for about 200 orbits.
During that time it behaves exactly like the other simulations with the exception that the width of the BL increases
much more quickly. This is in good agreement with the fact that the growth rate of the sonic instability depends on the
resolution in such a way that it is  faster when the  the resolution is higher (see Sect.~\ref{sec:instability}).
Since the BL widens to a large extent, it extends very far into the disk and into the star. The latter means that a
consistent simultaneous treatment of the star is required for numerical simulations involving large simulation times.
We will discuss this point in detail in the next  section.

\subsection{The long-term evolution of our simulations}\label{sec:long_term}

We ran and monitored our reference simulation for over 2000 orbits (see Fig.~\ref{fig:Ekin_of_t}), and the instability
and the periodic outbursts do not die out during that time. There is, however, a change in the pattern of
the acoustic waves that occurs around orbit 1200, which  coincides precisely with the moment when the 
base of the BL reaches the inner boundary of the simulation domain. We consider this a problematic moment, since it is not possible to treat
the star correctly in 2D $r$-$\varphi$ simulations. The geometry of the coordinate system inherently
gives the star a cylindrical shape in our simulations. In Sect.~\ref{sec:bcs} we  describe how we treat the
inner boundary in order to approximate the beginning of the star and that the radial velocity is set to outflow with
a certain velocity at this point. Therefore, we might expect unphysical and artificial processes once the vortices
of the acoustic modes reach this point because the radial velocity is fixed to a certain value and an inflow in the
domain is not supported. Furthermore, we force the azimuthal velocity to zero at the inner boundary. However, the
acoustic modes seem to be quite robust since they do not die out despite the difficulties with the artificial inner 
boundary. The change in the pattern at orbit 1200 might be due to a switching of the modes to the upper branch, since
the spiral arms of the waves are unwinding considerably and changing from leading to trailing. Another 1000 orbits
later the process is reversed again and the system switches back to the lower branch.

However, in order to capture the physics correctly at these late stages in time and to still make reliable statements
once the BL extends all the way to the surface of the star or even into the star, it is necessary to attach a realistic
model of the star to the disk. We consider, for instance, a wave launched in the BL that propagates toward the star. It can
 penetrate deep into the star before it is stopped by the steep density gradient and dissipates the angular momentum
somewhere in the star. On the other hand,  many problems arise when the star is treated consistently. The 
problem must then  be simulated in spherical coordinates, i.e. 3D simulations are necessary, and the
density  varies considerably from the interior of the star to the disk. Simulations of this kind are very demanding from
a computational point of view, but also very tempting, and we plan to investigate this problem in the future.

\section{Summary and conclusion}

We have performed highly resolved, 2D hydrodynamical simulations of the BL surrounding a young star in order to investigate the AM 
transport driven by instabilities. Extending previous simulations, we have a net mass flow through the disk,  we utilized a 
realistic equation of state, and (within the framework of one-temperature approximation) we included full radiative transfer 
in the disk plane as well as vertical cooling through an realistic estimate of the vertical optical depth, thereby
employing a quasi 3D radiation transport. Moreover, those simulations were started from state-of-the-art 1D
models of the BL and thus comprise realistic profiles for the density and temperature. We ran the simulations for
over two-thousand orbits in order to study the long-term behavior.

Our simulations confirm that the supersonic velocity drop in the BL is indeed prone to the sonic instability, a kind
of supersonic shear layer instability whose subsonic counterpart is the Kelvin-Helmholty instability \citep{2012ApJ...752..115B}.
Shortly after starting the simulations, the sonic instability sets in and
starts to grow rapidly for about 15 orbits. It then saturates and the BL is dominated by acoustic waves that propagate
both into the disk and into the star. These acoustic modes manifest themselves in  three distinct patterns
that can be related to the three branches of the instability of a plane parallel vortex sheet and do not die out for
the whole simulation time.

Additionally, the system repeatedly undergoes outbursts where the wave activity as well as the AM and mass transport
 increase considerably. We argue that these outburst are likely triggered by a secondary KH instability that develops in the
flat region of the azimuthal velocity profile due to several changes of sign of the vorticity. Since  the effective
temperature also shows strong variations during outbursts, they may play an
important role in explaining variations in the light curve, such as FU Or-outbursts or DNOs and QPOs in the case of BLs 
around white dwarfs. It is tempting to associate these phenomena with the outbursts
we observe in our simulations; however, this topic must be looked into further in order to draw reliable
conclusions. Such an investigation would certainly involve the study of how the outburst reacts to a change in parameters, 
among other things. It is also vital to perform 3D simulations on this question since there might be a dependence
on dimensionality \citep{2013ApJ...770...67B}.

Our main objective in this work was the clarification of the AM transport in unmagnetized astrophysical BLs.
We found that the acoustic modes indeed transport AM through the BL and the disk, and that this mechanism is even highly
efficient, reaching values for the dimensionless Reynolds stress $\alpha_\text{Re}$ of $\sim0.01$ in quiet states and
up to unity in high activity states (see Fig.~\ref{fig:alpha_of_t}). Along with AM, mass is also transported efficiently
through the domain (see Fig.~\ref{fig:Mdot_Jdot_of_t}). The BL reacts to the enhanced
transport by widening considerably until it reaches a thickness of about $0.2R_\ast$ (see Fig.~\ref{fig:width_bl}).

One of the most important elements of wave mediated AM transport is the intrinsic non-locality of this process, i.e. waves can
extract AM from one point in the disk and release it in another farther off. There are several implications that emerge from this:
Thus far, $\alpha$-models for the parametrization of the viscosity have been used both in the disk and in the BL,
possibly with some corrections for the BL. This prescription is justified in the disk where, under certain conditions, 
the system is susceptible to the MRI. The MRI leads to turbulence in the gas and angular
momentum is transported via turbulent stresses. This is a local process that also depends on the shearing at this point
and a local effective viscosity can be derived by the model of \citet{1973A&A....24..337S}. Furthermore, it allows for
efficient mixing of the gas through the turbulence. In contrast, the AM transport we have investigated in this work does
not depend on the local shearing, for instance, but is a non-local mechanism, instead. Therefore, it is doubtful whether
a conventional $\alpha$-model is applicable in the BL, and it will be hard to give a simple parametrization for
the AM transport in the BL at all. However, $\alpha_\text{Re}$ features a rather smooth behavior and might be
utilized for a simplified picture of the stresses in the BL. In addition, the mixing in the BL might  not be as
pronounced as in the disk, which could have important implications concerning the spectrum of the BL.

Connected with the notion of non-local AM transport is the fact that  the kinetic energy, which resides in the gas
shortly before it is decelerated to stellar rotation, does not need to be released as locally as  was previously assumed.
Waves  also transport energy and release it to the fluid where they are damped or dissipate. This might have important
implications for the structure and observational appearance of the BL. We consider, for instance, the alternative BL
theory of the \emph{spreading layer} \citep{1999AstL...25..269I,2010AstL...36..848I} where the gas is not decelerated
in the disk midplane but rather in two belts in the northern and southern hemisphere of the star. It is difficult to
bring a local viscosity model into accordance with this theory, since the friction on the surface of the neutron star
is far too small. The non-local AM transport described in this work could, however, shed new light upon the theory of
the spreading layer.

Simulations that investigate both the vertical structure and the non-axisymmetric properties can only be performed in
3D, however. This is a vital extension that should be taken as the next step in order to bring the models closer to
reality. In a 3D spherical geometry, it would also be possible to simultaneously model the star, a point that has
been illustrated in Sect.~\ref{sec:long_term}. Another major extension is the inclusion of magnetic fields, i.e. 
performing magnetohydrodynamical simulations of the BL. It would then be possible to relinquish any viscosity
prescription whatsoever and directly simulate the MRI in the disk and the acoustic modes in the BL. There might be
important interactions between the turbulent disc and the BL modes and new wave branches might add to the three discussed
here. However, the extensions mentioned above will greatly increase the computation time and possibly require  more
efficient codes and faster hardware.

\begin{acknowledgements}
Marius Hertfelder received financial support from the German National Academic Foundation (Studienstiftung des deutschen Volkes).
This research made use of \texttt{matplotlib} \citep{Hunter:2007}, a python module for data visualization.
Most of the simulations were performed on the bwGRiD cluster in Tübingen, which is funded by the Ministry for Education 
and Research of Germany and the Ministry for Science, Research and Arts of the state Baden-Württemberg, and the cluster
of the Collaborative Research Group FOR 759, funded by the DFG.
We also thank the referee for his constructive comments which helped to improve this paper.
\end{acknowledgements}

\bibliographystyle{aa}
\bibliography{aa26005-15}
\end{document}